\documentclass[a4paper,10pt]{article}

\usepackage{graphicx}
\usepackage{amssymb}
\usepackage{unit}
\usepackage{mathrsfs}
\usepackage{amsmath}
\usepackage{subfigure}
\usepackage{color}
\usepackage{amsthm}
\usepackage{amscd}
\usepackage{cite}

\def\iO{{\mathit{\Omega}}}
\def\bU{{\mathbf{U}}}
\def\bP{{\mathbf{P}}}
\def\bbR{{\mathbb{R}}}
\def\bbZ{{\mathbb{Z}}}
\def\bbX{{\mathbb{X}}}
\def\bbC{{\mathbb{C}}}
\def\cF{{\mathcal{F}}}

\title{%
Applied Koopman Operator Theory\\
for Power Systems Technology
}%

\author{%
Yoshihiko Susuki\footnote{Yoshihiko Susuki is with Department of Electrical and Information Systems, Osaka Prefecture University, Japan. \texttt{susuki@eis.osakafu-u.ac.jp}, \texttt{susuki@ieee.org}}~~~
Igor Mezi\'c\footnote{Igor Mezi\'c is with Department of Mechanical Engineering, University of California, Santa Barbara, United States. \texttt{mezic@engineering.ucsb.edu}}~~~
Fredrik Raak~~~Takashi Hikihara\footnote{Fredrik Raak and Takashi Hikihara are with Department of Electrical Engineering, Kyoto University, Japan.}
}%
\begin{document}

\maketitle

\begin{abstract}

Koopman operator is a composition operator defined for a dynamical system described by nonlinear differential or difference equation.  
Although the original system is nonlinear and evolves on a finite-dimensional state space, the Koopman operator itself is linear but infinite-dimensional (evolves on a function space).  
This linear operator captures the full information of the dynamics described by the original nonlinear system.  
In particular, spectral properties of the Koopman operator play a crucial role in analyzing the original system.  
In the first part of this paper, we review the so-called Koopman operator theory for nonlinear dynamical systems, with emphasis on modal decomposition and computation that are direct to wide applications.  
Then, in the second part, we present a series of applications of the Koopman operator theory to power systems technology.  
The applications are established as data-centric methods, namely, how to use massive quantities of data obtained numerically and experimentally, through spectral analysis of the Koopman operator:  coherency identification of swings in coupled synchronous generators, precursor diagnostic of instabilities in the coupled swing dynamics, and stability assessment of power systems without any use of mathematical models.   
Future problems of this research direction are identified in the last concluding part of this paper.

\end{abstract}



\section{Introduction}
\label{sec:intro}

Complex phenomena of the electric power system occur on a wide range of scales in both space and time.  
The power system has a more than 100 years of history in operation and is well-known as a large-scale, man-made, networked system of heterogeneous components with internal dynamics.  
The dynamic components include generation units (plants), loads, transmission lines, and---as a current particular interest---distributed renewable resources such as solar, wind, geothermal, and cogeneration plants.  
Examples of the complex phenomena are synchronization of individual rotating machines, voltage dynamics and collapse, and cascading outages leading to wide-area blackouts.  
Cascading outages have been recorded in practical power systems  \cite{Kurita_CDC88,Taylor_IEEECAP10,Kosterev_IEEETPS14,Sweden:2003,NorthAmerica:2003,Italy:2003,UCTE:2006,ArizonaSoCal:2011}.  
The details of the phenomena have been studied with direct numerical simulations of nonlinear mathematical models: see, e.g., \cite{Kosterev_IEEETPS14,Hiskens_IEEETPWRS14,Mani_BSDC-VI-2004,Kinney_EurPhysJ46,Dobson_CHAOS17,Susuki_IEICETEA09}.  
Analyzing such complex phenomena is of basic importance for realizing the so-called ``smart" society in the future.  

In this paper, we have a modest but counterintuitive goal:  to develop methodology and tools for analysis of such complex nonlinear power systems through \emph{linear} techniques.  
Needless to say, it has been demonstrated that a low-dimensional nonlinear dynamical system exhibits very complicated motions such as chaotic oscillation \cite{Ueda_CSF1}.  
For power system applications, such low-dimensional models play a crucial role in revealing  the key mechanisms of dynamic phenomena and instabilities: see, e.g., \cite{Abed_EPES6,Dobson_SCL13,Chiang_IEEETPWRS8,Ueda_IJBC8}.  
This direction of analyzing the nonlinear models is based on the \emph{state} description, called the \emph{Poincar\'e picture of dynamics}, and has great success in nonlinear dynamical systems  \cite{Guckenheimer:1983,Wiggins:1990} and nonlinear control systems \cite{Nijmeijier:1990}.  
However, it has been recognized that a high-dimensional nonlinear model with a huge number of states can not be easily handled and understood.  
To resolve the difficulty, as a counter-part of this description, B. O. Koopman is regarded as the pioneer for the use of linear transformations on Hilbert space to analyze Hamiltonian systems by introducing the so-called \emph{Koopman operator} and studying its spectrum \cite{Koopman_PNAS17}:  see \cite{Peterson_ET,Lasota_CFN} in detail.  
Even if the governing system is nonlinear and finite-dimensional, the Koopman operator is \emph{linear} but \emph{infinite-dimensional} and does not rely on linearization:  indeed, it captures the full information of the nonlinear dynamical system.  
In this setting, the elegant set of linear algebra, function space, and operator theory can be used to analyze the nonlinear system while avoiding the handling of a huge number of states.  
In addition to the linear treatment, the \emph{Koopman picture of dynamics} has a close connection to data-centric requirement to analysis of complex dynamics \cite{Marko_CHAOS22,Mezic_ARFM45}, where multiple domains---dynamical systems theory, large-scale simulation, sensing, data mining, and so on---can work together.  
This is currently desirable in power systems technology according to the new development of real-time Phasor Measurement Units (PMUs), which offers an advanced data collection method using phasors of AC (Alternative-Current) voltages: see, e.g., \cite{Phadke_IEEECAP6,Murphy_IEEECAP35,DeLaRee_IEEETSG1}. 

The purposes of this paper are two-fold.  
One is to provide a brief review of the Koopman operator theory of nonlinear dynamical systems. 
There is a large set of literature of theoretical progress in the Koopman operator: see, e.g., \cite{Mezic_PD197,Mezic_ND41,Marko_PD241,Alex_CHAOS22,Mauroy_PD261,Lan_PD242,Alex_CDC13,Ryan_Preprint,Mezic_CDC15,Hassan_Preprint} and \cite{Marko_CHAOS22,Mezic_ARFM45} for comprehensive reviews.  
In particular, it is shown in \cite{Mezic_ND41} via spectral analysis of the Koopman operator that single-frequency modes can be embedded in complex spatiotemporal dynamics.  
These modes are later named the \emph{Koopman Modes} (KMs) \cite{Rowley_JFM641} and used for characterizing global behaviors of complex fluid flows: see \cite{Rowley_JFM641,Phan:2015} and references of \cite{Mezic_ARFM45}.  
Due to high-dimensional, spatiotemporal nature of the complex phenomena in the power system, it is of basic interest for practitioners to identify a small number of dominant components or modes that approximates the phenomena observed practically and numerically.  
In the first part of this paper, we review a recent progress of the Koopman operator theory with emphasis on modal decomposition and \emph{computation}, which we believe are direct to wide applications beyond power systems technology addressed in this paper.  
The first part of this paper is generally described and not specific to any application domain.  

The other purpose is to present a series of applications of the Koopman operator theory to power systems technology that we have recently published since 2010.  
The applications are established as \emph{data-centric methods}, namely, how to use massive quantities of data obtained numerically and experimentally:  coherency identification of swings in coupled synchronous generators \cite{Susuki_IEEETPWRS26}, precursor diagnostic of instabilities in the coupled swing dynamics \cite{Susuki_IEEETPWRS27}, and stability assessment without any use of models \cite{Susuki_IEEETPWRS29}.  
It should be noted that other applications in power and energy systems have been reported: for example, network partitioning related to avoidance of cascading failures \cite{Raak_IEEETPWRS2015,Raak_IFACCPES15}, analysis of PMU data in Japan \cite{Ota_IEEJ2015}, analysis of building dynamics \cite{Eisenhower_SimBuild2012,Georgescu_EB86}, quantification of smoothing effects in wind energy \cite{Raak_IEEJ2016}.  
Outside the fluid mechanics and power systems technology, the Koopman operator theory has been applied to background/foreground separation in video \cite{Grosek_Preprint}, data fusion \cite{Matt_EPL109}, dynamic texture \cite{Surana_ACC15}, and so on.  
We will try to describe these applications so that readers can easily understand how the \emph{data-centric power system analysis} is developed via spectral properties of the Koopman operator.  

The rest of this paper is organized as follows:  
In Section~\ref{sec:theory} we provide a summarized theory of the Koopman operator for nonlinear dynamical systems based on the literature mentioned above.  
In Sections~\ref{sec:coherency} to \ref{sec:stability} we present a series of applications of the Koopman operator theory in power systems technology based on our previous papers \cite{Susuki_IEEETPWRS26,Susuki_IEEETPWRS27,Susuki_IEEETPWRS29}. 
Conclusion of this paper is made in Section~\ref{sec:outro} with future problems.  

\textbf{Notation}~~~All vectors are viewed as columns.  
The sets of all real numbers, complex numbers, and integers are denoted by $\bbR$, $\bbC$, and $\bbZ$, respectively. 
We use $\mathbb{R}_{\geq 0}:=\{x : x\in\bbR, x\geq 0\}$.  
The symbol $\mathbb{T}$ represents the one-torus or circle.  
The symbol $\ii$ stands for the imaginary number, $|z|$ for the modulus of $z\in\bbC$, $\mathrm{Arg}(z)$ for its argument, and $z^\ast$ for the complex conjugate.  
For a vector $\vct{z}\in\bbC^m$ ($m$ times), $\vct{z}^\top$ stands for its transpose, $\vct{z}^\ast$ for the complex-conjugate transpose, and $||\vct{z}||:=\sqrt{\vct{z}^\ast\vct{z}}$ for the $\bbC^m$-norm.  
We use $\vct{z}^\mathrm{c}:=(\vct{z}^\ast)^\top$.   
The symbols of transpose and complex-conjugate transpose are also used for matrices.  
For a matrix $\mathsf{A}$, its rank is denoted as $\mathrm{rank}(\mathsf{A})$, and the linear space spanned by all linear combinations of the column vectors in $\mathsf{A}$ as $\mathrm{span}\{\mathsf{A}\}$.  
We use $\mathrm{diag}(a_1,a_2,\ldots)$ for a diagonal matrix whose diagonal entries starting in the upper left corner are $a_1,a_2,...$
The symbol $\bot$ indicates the orthogonality between vector and space.

\section{Summarized theory of Koopman operators}
\label{sec:theory}

This section provides a summarized theory of Koopman operators for nonlinear dynamical systems.  
Comprehensive reviews of the emerging theory exist in \cite{Marko_CHAOS22,Mezic_ARFM45}.  
Current emphasis is on \emph{computation} as a tool for how the theory is utilized for solving concrete dynamical problems, 
which we believe is direct to wide applications beyond power systems.

\subsection{Definition}
\label{subsec:def}

Throughout the paper, we consider a dynamical system evolving on a finite $n$-dimensional manifold $\bbX$:  for continuous time $t\in\bbR$,
\begin{equation}
\frac{d}{dt}\vct{x}(t)=\vct{F}(\vct{x}(t)),
\label{eqn:cds}
\end{equation}
and for discrete time $k\in\bbZ$,
\begin{equation}
\vct{x}[k+1]=\vct{T}(\vct{x}[k]),
\label{eqn:dds}
\end{equation}  
where $\vct{x}$ is called the \emph{state} belonging to the \emph{state space} $\bbX$.  
The function $\vct{F}: \bbX\to\mathrm{T}\mathbb{X}$ (tangent bundle of $\mathbb{X}$) is a nonlinear vector-valued function and is assumed to be tractable in the region we are interested, and $\vct{T}: \bbX\to\bbX$ a nonlinear vector-valued map assumed in the same manner.  
For the continuous-time system (\ref{eqn:cds}), the following finite time-$h$ map $\vct{S}_{t\to t+h}$ ($h\in\bbR$) is defined as
\[
\vct{S}_{t\to t+h}: \bbX\to\bbX;~~\vct{x}(t)\mapsto\vct{x}(t+h)=\vct{x}(t)
+\int^{t+h}_t\vct{F}(\vct{x}(\tau))d\tau.
\]
Since (\ref{eqn:cds}) is autonomous, the map does not depend on $t$, simply denoted as $\vct{S}_h$.  
The one-parameter group of maps, $\{\vct{S}_h; h\in\bbR\}$, is called the \emph{flow}.  
Below, we will use $t$ for the suffix of the flow instead of $h$ for clarity of the presentation.  

Here we introduce the Koopman operators for the continuous- and discrete-time dynamical systems.  
To do this, the so-called \emph{observable} $f$ is introduced as a scalar-valued function defined on the state space $\bbX$, namely
\[
f: \bbX\to\bbC.
\]
This observable is a mathematical formulation of observation of the state's dynamics and is well-known as the \emph{output} function in the control engineering community.  
Below, we will denote by $\cF$ a given space of observables (its properties will be mentioned later).  
For an observable $f\in\cF$, the \emph{Koopman operator} $\bU_t$ for the continuous-time system (\ref{eqn:cds}) and $\bU$ for the discrete-time system (\ref{eqn:dds}) map $f$ into a new function as follows:
\begin{eqnarray}
\bU_tf &:=& f\circ\vct{S}_t,
\nonumber\\ 
\bU f &:=& f\circ\vct{T}.
\nonumber 
\end{eqnarray}
That is, the Koopman operator $\bU_t$ (or $\bU$) describes the time $t$ (or $k$) evolution of observation (output of the system) along the state's dynamics:  for the continuous-time case, the output $y(t)$ is described as
\begin{equation}
y(t):=f(\vct{x}(t))=f(\vct{S}_t(\vct{x}(0)))
=f\circ\vct{S}_t(\vct{x}(0))
=(\bU_tf)(\vct{x}(0)).
\label{eqn:output_cds}
\end{equation}
Although the dynamical systems (\ref{eqn:cds}) and (\ref{eqn:dds}) can be nonlinear and evolve in the finite-dimensional space, the Koopman operators are \emph{linear} but \emph{infinite-dimensional}.  
The linearity is easily proven: for the continuous-time case, by picking up any two observables $f_1,f_2\in\cF$ and two scalars $\alpha_1,\alpha_2\in\bbC$, we have
\[
\bU_t(\alpha_1f_1+\alpha_2f_2)=(\alpha_1f_1+\alpha_2f_2)\circ\vct{S}_t
=\alpha_1(f_1\circ\vct{S}_t)+\alpha_2(f_2\circ\vct{S}_t)
=\alpha_1(\bU_tf_1)+\alpha_2(\bU_tf_2).
\]
This type of composition operator is defined for a large class of nonlinear dynamical systems \cite{Lasota_CFN} and does not rely on linearization: indeed, it captures the full information on the original nonlinear systems.  
The essential idea outlined in this paper is clear: to analyze dynamics described by the nonlinear systems (\ref{eqn:cds}) and (\ref{eqn:dds}) through the linear operators.

Here it is noted which class of function space is chosen for $\cF$ of observables.  
In the traditional ergodic theory, which has mainly studied measure-preserving dynamical systems on a compact space, the standard $L^2$ space is chosen.  
For this space, the celebrated Birkhoff's point-wise ergodic theorem and so on are presented: see \cite{Arnold:1968} in detail.     
For a general, possibly dissipative dynamical system on a non-compact space, there is still room for consideration.  
Even in the dissipative system, if it has a compact attractor, the $L^2$ space is effectively used for investigating asymptotic dynamics on the attractor through the Koopman operator.  
In \cite{Mauroy_PD261}, for a nonlinear system with a stable equilibrium point, the space of analytic functions in the domain of attraction of the point is chosen as $\cF$.  
Also, in \cite{Ryan_Preprint} generalized Hardy space is introduced as $\cF$ for analysis of off-attractor dynamics of a nonlinear system with a stable equilibrium point or limit cycle.  
As a summary, the choice of a space of analytic functions is better for analyzing dissipative dynamics via the Koopman operator with a relatively-simple spectrum.\footnote{except for a nonlinear system with a chaotic attractor}

\subsection{Spectral properties}
\label{subsec:spectral}

Spectrum plays an important role in understanding the Koopman operators and, in our purpose, investigating the original nonlinear systems.  
Since the linear operators are infinite-dimensional, it is necessary to consider both discrete and continuous spectra \cite{Kato,Yoshida}.  

First of all, let us introduce the discrete part of spectrum, that is, \emph{eigenvalues} of the Koopman operators.   
In this part, the concept in finite-dimensional matrices is naturally applied.  
For the continuous-time system (\ref{eqn:cds}), eigenfunctions and eigenvalues of $\bU_t$ are defined as follows: for non-zero $\phi_j\in\cF$ and $\nu_j\in\bbC$ such that 
\begin{equation}
\bU_t\phi_j=\ee^{\nu_jt}\phi_j,\makebox[2em]{}
j=1,2,\ldots
\label{eqn:ds_cds}
\end{equation}
The functions $\phi_j$ are referred to as \emph{Koopman eigenfuctions}, and the constants $\nu_j$ as the associated \emph{Koopman Eigenvalues} (KEs).  
The set of all the KEs is called the \emph{discrete spectrum} of  $\bU_t$.  
Also, for the discrete-time system (\ref{eqn:dds}), the definitions of Koopman eigenfunctions and eigenvalues are almost same:  for non-zero $\phi_j\in\cF$ and $\lambda_j\in\bbC$ such that 
\[
\bU\phi_j=\lambda_j\phi_j,\makebox[2em]{}
j=1,2,\ldots
\]
For example, consider the following first-order continuous-time system: for $\theta\in\mathbb{T}$,
\begin{equation}
\frac{d}{dt}\theta(t)=\iO,\makebox[2em]{}
\iO:\textrm{real constant}.
\label{eqn:ex_cds}
\end{equation}
The corresponding flow is $S_t(\theta)=\theta+\iO t~(\textrm{mod}\,2\pi)$.  
It is easily checked that the $j$-th Fourier basis $\phi_j(\theta)=\ee^{\ii j\theta}$ satisfies (\ref{eqn:ds_cds}) with KE $\nu_j=\ii j\iO$:
\[
(\bU_t\phi_j)(\theta)
=\phi_j(S_t(\theta))
=\ee^{\ii j(\theta+\iO t)}
=\ee^{\ii j\iO t}\ee^{\ii j\theta}
=\ee^{\ii j\iO t}\phi_j(\theta).
\]
For a suitable choice of $\cF$, we can expand the observable $f(\theta)$ in terms of the Fourier bases and obtain its composition with $S_t$:
\begin{equation}
(\bU_tf)(\theta)
=\sum_{j\in\bbZ}c_j(\bU_t\phi_j)(\theta)
=\sum_{j\in\bbZ}c_j\phi_j(\theta)\ee^{\ii j\iO t},
\label{eqn:linear_cds}
\end{equation}
where $c_j\in\bbC$ is a coefficient for the expansion.  
The formula (\ref{eqn:linear_cds}) suggests that the time evolution of observable under (\ref{eqn:ex_cds}) is expressed as an infinite \emph{summation over $\mathbb{Z}$} in oscillations with \emph{distinct}, \emph{constant} (angular) frequencies $j\iO$ of KEs.

Next, we mention the continuous part of spectrum that is of particular interest in infinite-dimensional linear operators.  
The mathematical definition of $\emph{continuous spectrum}$ is available in textbooks (see, e.g., \cite{Kato,Yoshida}), and we now describe its dynamical motivation with the simple, second-order continuous-time linear system: for $(\theta,\omega)\in\mathbb{T}\times\bbR$,
\[
\frac{d}{dt}\theta(t)=\omega(t),\makebox[2em]{}
\frac{d}{dt}\omega(t)=0.
\]
The induced flow is $\vct{S}_t(\theta,\omega)=[\theta+\omega t\,(\textrm{mod}\,2\pi),\omega]^\top$.  
Obviously, the value $0$ corresponds to an eigenvalue with eigenfunction $\phi_0(\theta,\omega)=\omega$, which can be replaced by any function of $\omega$. 
Here, since the state space is periodic in $\theta$, in the similar manner as above, we may expand the observable $f(\theta,\omega)$ in terms of the Fourier bases like
\[
f(\theta,\omega)=\omega+\sum_{j\in\bbZ\setminus\{0\}}a_j(\omega)\ee^{\ii j\theta},
\] 
where $a_j(\omega)$ is a function dependent on $f$.  
Thus, we have the following composition of $f$ with $\vct{S}_t$:
\[
(\bU_tf)(\theta,\omega)
=\omega+\sum_{j\in\bbZ\setminus\{0\}}a_j(\omega)\ee^{\ii j(\theta+\omega t)}
=\omega+\sum_{j\in\bbZ\setminus\{0\}}\left\{a_j(\omega)\ee^{\ii j\theta}\right\}\ee^{\ii j\omega t}.
\]
The formula is apparently similar to (\ref{eqn:linear_cds}), but the angular frequencies depend on the state $\omega$ that can be taken in $\bbR$.  
This suggests that no decomposition of the time evolution $f(\theta(t),\omega(t))$ based on distinct, constant frequencies is obtained.  
Rather than this, the time evolution is represented in terms of \emph{continuous} quantity---\emph{integral over} $\bbR$.  
This is a situation in which the continuous spectrum appears in the Koopman operator.  

Now, based on \cite{Mezic_ND41}, we introduce the spectral representation of the Koopman operator $\bU$ that is unitary on $\cF=L^2(\bbX)$ with a compact $\bbX$.  
The unitary case has been traditionally studied in ergodic theory \cite{Peterson_ET} because it results from a measure-preserving map on a compact state space.  
The following representation holds for the same as for the continuous-time case. 
For unitary $\bU$, its spectrum is restricted to the unit circle in the complex domain.  
Thus, $\bU$ admits a unique decomposition into its singular and regular parts:
\begin{equation}
\bU=\bU\sub{s}+\bU\sub{r},
\label{eqn:sd-unitary}
\end{equation}
where there exists a (Hilbert) subspace $\mathcal{H}\subset\cF$ such that for any $f\in\cF$, the new function $\bU\sub{s}f$ belongs in $\mathcal{H}$, and $\bU\sub{r}f$ does in the orthogonal complement of $\mathcal{H}$ .  
The singular part $\bU\sub{s}$ implies the discrete spectrum, determined by the KEs of $\bU$:
\[
\bU\sub{s}=\sum^{\infty}_{j=1}\lambda_j\bP_j,
\]
where $\bP_j$ is the projection operator associated with KE $\lambda_j$ of unity modulus and eigenfunction $\phi_j$; $\bP_i\circ\bP_j=\delta_{ij}\bP_i$ where $\delta_{ij}$ is the Kronecker delta.  
The regular part $\bU\sub{r}$ implies the continuous spectrum and is represented with a continuous spectral measure $E(\theta)$ on $\mathbb{T}$ as
\[
\bU\sub{r}=\int_\mathbb{T}\ee^{\ii \theta}dE(\theta).
\]

Last of all, in terms of data-centric applications, we introduce a connection of discrete spectra between the continuous- and discrete-time systems.  
For this, consider a uniformly-sampled observation by letting $T(>0)$ be the sampling period.  
From (\ref{eqn:output_cds}) we have
\[
y[k]:=y(kT)=(\bU_{kT}f)(\vct{x}(0))=((\bU_T)^kf)(\vct{x}(0)).
\]
This clearly shows that the sampled observation of the dynamics described by the continuous-time system (\ref{eqn:cds}) is modeled by the Koopman operator $\bU=\bU_T$ for the discrete-time system (\ref{eqn:dds}) with eigenfunctions $\phi_j$ and eigenvalues $\lambda_j=\ee^{\nu_jT}$.   
Note that the Koopman eigenfunctions $\phi_j$ are common for the original continuous- and derived discrete-time systems.  
Since the main idea in this paper is to analyze the dynamics from sampled observational data, we mainly focus on the discrete-time setting in the following.

\subsection{Koopman modes: formulation and computation}
\label{subsec:KM}

\subsubsection{Mathematical formulation}
\label{subsec:KM-math}

The idea in \cite{Mezic_ND41} is to analyze dynamics governed by the nonlinear system (\ref{eqn:dds}) using spectrum of the Koopman operator.   
First of all, let us consider the time evolution of observable, $f(\vct{x}[k])$, for the unitary case of $\bU$.  
Now, from its spectral decomposition in (\ref{eqn:sd-unitary}), we have
\[
f(\vct{x}[k])=(\bU^kf)(\vct{x}[0])
=(\bU^k\sub{s}f+\bU^k\sub{r}f)(\vct{x}[0])
=\sum_{j=1}^\infty\lambda^k_j(\bP_jf)(\vct{x}[0])
+\int_\mathbb{T}\ee^{\ii k\theta}dE(\theta)f(\vct{x}[0]),
\]
where we used the property of the projection operator $\bP_j$.  
By using the formula $\bP_jf=\phi_jV_j$, where $V_j\in\bbC$ is a constant for the projection depending on $f$, we obtain the following spectral representation of $f(\vct{x}[k])$ in terms of $\bU$:
\[
f(\vct{x}[k])
=\sum_{j=1}^\infty\lambda^k_j\phi_j(\vct{x}[0])V_j
+\int_\mathbb{T}\ee^{\ii k\theta}dE(\theta)f(\vct{x}[0]).
\]
The first term represents the contribution of KEs to $\{\vct{f}(\vct{x}[k])\}$ and describes the average and quasi-periodic parts in $\{\vct{f}(\vct{x}[k])\}$.  
The second term represents the contribution of continuous spectrum and describes the aperiodic part contained in $\{\vct{f}(\vct{x}[k])\}$.  

Furthermore, we can extend this representation for a \emph{vector-valued} observable $\vct{f}:=[f_1,\ldots,f_m]^\top: \bbX\to\bbC^m$, where $f_i\in\cF$.  
The observable corresponds to a vector of any quantities of interest such as frequencies and voltages metered at various points in a power system, which is direct to the applications in Sections~\ref{sec:coherency} to \ref{sec:stability}.  
Then, by the similar argument to the scalar-valued observable, the time evolution $\vct{f}(\vct{x}[k])=[f_1(\vct{x}[k]),\ldots,f_m(\vct{x}[k])]^\top$ is exactly represented as follows:
\begin{equation}
\vct{f}(\vct{x}[k])=\sum_{j=1}^{\infty}\lambda^k_j\phi_j(\vct{x}[0])\vct{V}_j+
\left[
\begin{array}{c}
{\displaystyle \int_\mathbb{T}\ee^{\ii k\theta}dE(\theta)f_1(\vct{x}[0])}\\
\vdots\\
{\displaystyle \int_\mathbb{T}\ee^{\ii k\theta}dE(\theta)f_m(\vct{x}[0])}
\end{array}
\right].
\label{eqn:sr-vector}
\end{equation}
Thus, we refer to $\vct{V}_j\in\bbC^m$ as the \emph{Koopman Mode} (KM) (named in \cite{Rowley_JFM641}) associated with KE $\lambda_j$, corresponding to the observable $\vct{f}$.  
The KE $\lambda_j$ characterizes the temporal behavior of the corresponding KM $\vct{V}_j$: the argument of $\lambda_j$ determines its frequency.  
The representation (\ref{eqn:sr-vector}) shows via spectral analysis of the Koopman operator that single-frequency modes can be embedded in the complex spatio-temporal dynamics.

It is shown in \cite{Mezic_PD197,Mezic_ND41} that the terms $\phi_j(\vct{x}[0])\vct{V}_j$ are defined and computed with a projection operation associated with $\bU$ applied to the observable $\vct{f}$.  
The projection operation of a non-zero $\bP^\nu$ with KE $\lambda=\ee^{\ii 2\pi\nu}$ is constructively represented as follows: 
\[
\bP^\nu f(\vct{x}[0])
=\lim_{n\rightarrow\infty}\frac{1}{n}\sum^{n-1}_{k=0}\ee^{-\ii2\pi k\nu}f(\vct{x}[k]),\makebox[2em]{}
f\in\cF,
\]
where $\nu\in[-1/2,1/2)$.  
The projections of the $p$ components $f_1,\ldots,f_m$ of $\vct{f}$ based on $\bP_j$ are thus obtained:
\begin{equation}
\left[
\begin{array}{c}
\bP^{\nu_j}f_1(\vct{x}[0])\\
\vdots\\
\bP^{\nu_j}f_m(\vct{x}[0])
\end{array}
\right]
=\phi_j(\vct{x}[0])\vct{V}_{j},
\label{eqn:projection}
\end{equation}
where $\nu_j=\mathrm{Im}[\ln\lambda_j]/2\pi$.   
This formula (\ref{eqn:projection}) associates $\phi_j(\vct{x}[0])\vct{V}_j$ with the projection operation based on $\bP^\nu$.   
The left-hand sides of (\ref{eqn:projection}) are like the discrete Fourier transform of data $\{\vct{f}(\vct{x}[k])\}$, and hence $\phi_j(\vct{x}[0])\vct{V}_j$ can be directly computed from data.  

Until now, we have assumed that the Koopman operator is unitary, that is, a case where the dynamics of (\ref{eqn:dds}) are on an attractor.  
Even if this is not the case, that is, we consider dynamics off attractors of (\ref{eqn:dds}), each KM oscillates with a single frequency.   
If each of the $p$ components of $\vct{f}$ lies within the span of eigenfunctions $\phi_j$, then, as in \cite{Rowley_JFM641}, we may expand $\vct{f}$ in terms of the Koopman eigenfunctions as
\[
\vct{f}=\sum_{j=1}^{\infty}\phi_j\vct{V}_j, 
\]
where $\vct{V}_j$ are regarded as vector coefficients for the expansion.  
The time evolution $\{\vct{f}(\vct{x}[k])\}$ is identically given as
\begin{equation}
\vct{f}(\vct{x}[k])
=\sum^\infty_{j=1}\phi_j(\vct{x}[k])\vct{V}_j
=\sum^\infty_{j=1}{\bU}^k\phi_j(\vct{x}[0])\vct{V}_j
=\sum^\infty_{j=1}\lambda^k_j\phi_j(\vct{x}[0])\vct{V}_j.
\label{eqn:KMD}
\end{equation}
Also, if the dynamics observed here have only a finite number of peaks in the frequency domain (practical experience suggests that this situation is normal in power systems), then we can expect the expansion gives an approximation of the dynamics.  
For dynamics off attractors, the KE $\lambda_j$ still characterizes the temporal behavior of the corresponding KM $\vct{V}_j$:  
the magnitude of $\lambda_j$ determines the growth rate.  
We call the decomposition (\ref{eqn:KMD}) based on the discrete spectrum of the Koopman operator the \emph{Koopman Mode Decomposition} (KMD).
We will show its capability in power systems in the latter part of this paper, Sections~\ref{sec:coherency} to \ref{sec:stability}.  

\subsubsection{Algorithms for computation}
\label{subsec:KM-algo}

For data-centric applications, KMD is considered directly from time-series data without any use of description of the original system, such as explicit description of $\vct{T}$ and $\vct{F}$.  
A common idea for the existing KMD algorithms is to find a finite-dimensional approximation of the action of the Koopman operator directly from finite-time data.  
The following description in this paragraph is based on \cite{Marko_CHAOS22}.  
Now consider a fixed vector-valued observable $\vct{f}=[f_i,\ldots,f_m]^\top\in\cF^m$ ($m$ times) and associated finite-time data (snapshots) $\{\vct{y}_k=\vct{f}(\vct{x}[k]); k=0,\ldots,N-1, N<\infty\}$.  
Then, the following function space $\mathcal{K}_N$ is sufficient for investigation of the finite-time data:
\[
\mathcal{K}_N:=\mathrm{span}\{\mathbf{\underline{U}}^k\vct{f}; k=0,\ldots,N-1\},
\makebox[2em]{}\mathbf{\underline{U}}^k\vct{f}:=[\bU^kf_1,\ldots,\bU^kf_m]^\top.
\]
In the following, we will assume that $\{\mathbf{\underline{U}}^k\vct{f}; k=0,\ldots,N-1\}$ is a linearly independent set so that these functions form a basis for $\mathcal{K}_N$.  
Here, let $\bP_N: \cF^m\to\mathcal{K}_N$ be a projection from the space of vector-valued observables onto $\mathcal{K}_N$. 
Thus, the following operator
\[
\bP_N\mathbf{\underline{U}}: \mathcal{K}_N\to\mathcal{K}_N
\]
is finite-dimensional and has a matrix representation, $\mathsf{C}_N: \bbC^N\to\bbC^N$, in terms of the $\{\mathbf{\underline{U}}^k\vct{f}\}$-basis. 
Note that the matrix $\mathsf{C}_N$ is dependent on (i) vector-valued observable $\vct{f}$, (ii) the number $N$ of snapshots, and (iii) the projection $\bP_N$.  
If $(\lambda,\vct{V}_\lambda)$ is a pair of eigenvalue and eigenvector of $\mathsf{C}_N$, where $\vct{V}_\lambda=[V_{\lambda,0},\ldots,V_{\lambda,N-1}]^\top\in\bbC^N$, then $\lambda$ is an eigenvalue of $\bP_N\mathbf{\underline{U}}$ and $\vct{\phi}_\lambda:=\sum^{N-1}_{j=0}V_{\lambda,j}(\mathbf{\underline{U}}^j\vct{f})$ the associated eigenfunction.  
By restricting our attention to a fixed observable $\vct{f}$ and associated space $\mathcal{K}_N$, we have reduced the problem of locating eigenvalues and modes of the Koopman operator to locating eigenvalues and eigenvectors for a matrix $\mathsf{C}_N$.  
It is shown in \cite{Hassan_Preprint} that as the number $N$ of snapshots goes to infinity, the KMs located with $\bP_N\mathbf{\underline{U}}$ approach to exact KMs with respect to the fixed $\vct{f}$ for the Koopman operator $\mathbf{\underline{U}}$.  
The existing KMD algorithms provide a procedure of how to compute $\mathsf{C}_N$ directly from data.  
Below, we introduce three existing algorithms: Arnoldi-type algorithm \cite{Rowley_JFM641}, dynamic mode decomposition \cite{Schmid_JFM656,Chen_JNLS22,Jovanovic_PF26,Tu_JCD1}, and vector Prony analysis \cite{Susuki_CDC15}. 

\vspace*{2mm}
\noindent{\bf (a)~Arnoldi-type algorithm}

First, the Arnoldi-type algorithm \cite{Rowley_JFM641} for computation of KMs and KEs is introduced.  
Consider the finite-time data under uniform sampling, again given by
\begin{equation}
\{\vct{y}_0,\vct{y}_1,\ldots,\vct{y}_{N-1}\},
\label{eqn:data_N}
\end{equation}
where $\vct{y}_k\in\bbR^m$ is the $k$-th snapshot and $N$ the number of available snapshots. 
Then, the so-called \emph{empirical Ritz values} $\tilde{\lambda}_j$ and \emph{empirical Ritz vectors} $\vct{\tilde{V}}_j$ of the data are defined with the following algorithm:
\begin{itemize}
\item[(i)] Define a constant vector $\vct{c}=[c_0,\ldots,c_{N-2}]^\top$ such that for vector $\vct{r}\in\bbR^m$ satisfying $\vct{r}\bot$ $\mathrm{span}\{\mathsf{K}_{N-1}\}$,
\begin{equation}
\vct{r}
=\vct{y}_{N-1}-\mathsf{K}_{N-1}\vct{c},
\label{eqn:c}
\end{equation}
with the \emph{Krylov subspace} defined as
\begin{equation}
\mathsf{K}_{N-1}:=[\vct{y}_0,\ldots,\vct{y}_{N-2}].
\label{eqn:K_N-1}
\end{equation}
\item[(ii)] Define the matrix $\mathsf{C}_{N-1}$ as
\[
\mathsf{C}_{N-1}:=
\left[
\begin{array}{ccccc}
0 & 0 & \cdots & 0 & c_0\\
1 & 0 & \cdots & 0 & c_1\\
0 & 1 & \cdots & 0 & c_2\\
\vdots & \vdots & \ddots & \vdots & \vdots\\
0 & 0 & \cdots & 1 & c_{N-2}\\
\end{array}
\right],
\]
called the \emph{companion matrix}.  
Then, locate its $N-1$ eigenvalues as the empirical Ritz values $\tilde{\lambda}_1,\ldots,\tilde{\lambda}_{N-1}$.  
\item[(iii)] Define the \emph{Vandermonde} matrix $\mathsf{T}$ using $\tilde{\lambda}_j$ as
\begin{equation}
\mathsf{T}:=
\left[
\begin{array}{ccccc}
1 & \tilde{\lambda}_1 & \tilde{\lambda}^2_1 & \cdots  & \tilde{\lambda}^{N-2}_1\\
1 & \tilde{\lambda}_2 & \tilde{\lambda}^2_2 & \cdots  & \tilde{\lambda}^{N-2}_2 \\
\vdots & \vdots & \vdots & \ddots & \vdots\\
1 & \tilde{\lambda}_{N-1} & \tilde{\lambda}^2_{N-1} & \cdots & \tilde{\lambda}^{N-2}_{N-1} \\
\end{array}
\right].
\end{equation}
\item[(iv)] Define the empirical Ritz vectors $\vct{\tilde{V}}_j$ to be the columns of $\mathsf{V}:=\mathsf{K}_{N-1}\mathsf{T}^{-1}$.  
\end{itemize}
It is shown in \cite{Rowley_JFM641} that if all of the empirical Ritz values are non-zero and distinct, then the following decompositions of the data are obtained:
\begin{equation}
\vct{y}_k=\sum^{N-1}_{j=1}\tilde{\lambda}^k_j\vct{\tilde{V}}_j,\makebox[1em]{}
k=0,\ldots,N-1,\makebox[2em]{}
\vct{y}_{N-1}=\sum^{N-1}_{j=1}\tilde{\lambda}^{N-1}_j\vct{\tilde{V}}_j+\vct{r}.
\label{eqn:finite}
\end{equation}
Comparing with (\ref{eqn:KMD}), the empirical Ritz values $\tilde{\lambda}_j$ and vectors $\vct{\tilde{V}}_j$ behave precisely in the same manner as the KEs $\lambda_i$ and the terms $\phi_i(\vct{x}[0])\vct{V}_i$ containing the Koopman eigenfunctions and KMs, but for the finite sum (\ref{eqn:finite}) instead of the infinite sum (\ref{eqn:KMD}).  

Now, it is necessary to show how to choose the constant vector $\vct{c}$ from the data (\ref{eqn:data_N}).    
It is clearly stated in \cite{Marko_CHAOS22} that this is to minimize the $\bbC^{m}$-norm of $\vct{r}$ in (\ref{eqn:c}), corresponding to choosing the projection operator $\mathbf{P}_{N-1}$ so that $\mathbf{P}_{N-1}\mathbf{\underline{U}}^{N-1}\vct{f}$ is the least-squares approximation to $\mathbf{\underline{U}}^{N-1}\vct{f}$ at $\vct{x}_0\in\mathbb{X}$ satisfying $\vct{y}_0=\vct{f}(\vct{x}_0)$.  
Here, because of $\vct{r}_{N-1}\bot\,\mathrm{span}\{\mathsf{K}_{N-1}\}$, by multiplying $\mathsf{K}_{N-1}^\top$ on both sides of (\ref{eqn:c}) from left, we have the following linear equation:
\begin{equation}
\vct{0}=\mathsf{K}_{N-1}^\top\vct{y}_{N-1}-\left(\mathsf{K}_{N-1}^\top\mathsf{K}_{N-1}\right)\vct{c}.
\label{eqn:Ac=b}
\end{equation}
If $\mathsf{K}_{N-1}^\top\mathsf{K}_{N-1}\in\bbR^{(N-1)\times(N-1)}$ is regular, then there exists the unique solution $\vct{c}$ for (\ref{eqn:Ac=b}).  
If this is not the case, we use the Moore-Penrose pseudo-inverse.  
Especially, for data with low spatial dimension, that is, $m\ll N$ (this situation normally appears in power system applications), because of $\mathrm{rank}(\mathsf{K}_{N-1})\leq m$, the rank-deficiency of $\mathsf{K}_{N-1}$, namely the singularity of $\mathsf{K}_{N-1}^\top\mathsf{K}_{N-1}$ is inevitable when choosing $\vct{c}$.  
This implies no uniqueness of $\vct{c}$, and that the derived decomposition and information (frequencies, modes, etc.) should be cross-checked with other means of analysis such as the Fourier-based formula (\ref{eqn:projection}) and standard Fourier analysis.  
We will use the algorithm in Sections~\ref{sec:coherency} to \ref{sec:stability} and show the cross-check in Section~\ref{sec:coherency}.  

Here, based on \cite{Chen_JNLS22,Marko_CHAOS22}, let us see that the companion matrix $\mathsf{C}_{N-1}$ is a finite-dimensional approximation of the action of $\mathbf{\underline{U}}$.   
Using $\mathsf{C}_{N-1}$, although not mathematically rigorous, we have
\begin{eqnarray}
\mathbf{\underline{U}}[\vct{y}_0,\ldots,\vct{y}_{N-2}]
&=&[\vct{y}_1,\ldots,\vct{y}_{N-1}] \label{eqn:hoge}\\
&=&[\vct{y}_0,\ldots,\vct{y}_{N-2}]\mathsf{C}_{N-1}+\vct{r}\vct{e}^\top,\nonumber
\end{eqnarray}
where $\vct{e}^\top:=[0,\ldots,0,1]$. 
This suggests that if the last term $\vct{r}\vct{e}^\top$ is less dominant (e.g., $|\vct{r}|$ is close to zero), then $\mathsf{C}_{N-1}$ is thought of as an approximation to the action of the Koopman operator onto the finite-dimensional space $\mathcal{K}_{N-1}$.  
Based on the fact that the Vandermonde matrix diagonalizes the companion matrix as long as its eigenvalues are distinct, namely $\mathsf{C}_{N-1}=\mathsf{T}^{-1}\mathsf{\tilde{\Lambda}}\mathsf{T}$, where $\mathsf{\tilde{\Lambda}}:=\mathrm{diag}(\lambda_1,\ldots,\lambda_{N-1})$, we have
\begin{eqnarray}
\mathbf{\underline{U}}\mathsf{K}_{N-1}\mathsf{T}^{-1} 
&=& \mathsf{K}_{N-1}\mathsf{T}^{-1}\mathsf{\tilde{\Lambda}}+\vct{r}\vct{e}^\top\mathsf{T}^{-1}
\nonumber\\
\mathbf{\underline{U}}\mathsf{V}&=& \mathsf{V}\mathsf{\tilde{\Lambda}}+\vct{r}\vct{e}^\top\mathsf{T}^{-1}.
\nonumber
\end{eqnarray}
That is, if $||\vct{r}\vct{e}^\top\mathsf{T}^{-1}||$ is small, then $\mathsf{V}$ and $\mathsf{\tilde{\Lambda}}$ approximate eigenvectors and eigenvalues of $\mathbf{\underline{U}}$, respectively. 

\vspace*{2mm}
\noindent{\bf (b)~Dynamic mode decomposition}

Second, the Dynamic Mode Decomposition (DMD) is a well-known algorithm in the fluid mechanics community \cite{Schmid_JFM656,Jovanovic_PF26}.  
Connections with the aforementioned Arnoldi-type algorithm are studied in \cite{Chen_JNLS22,Tu_JCD1}.  
DMD is also applied in \cite{Barocio_IEEETPWRS30} to characterizing global behaviors of power systems. 
Also, an interesting extension of DMD for numerically approximating not only KEs and KMs but also Koopman eigenfunctions is proposed in \cite{Matt_JNLS25}.  

Since many comprehensive literatures on DMD exist, we here provide a brief introduction of DMD.  
The author of \cite{Schmid_JFM656} considers the form (\ref{eqn:K_N-1}) of a snapshot sequence.  
Here, let us consider a singular value decomposition of $\mathsf{K}_{N-1}$ as $\mathsf{Q}_1\mathsf{\Sigma}\mathsf{Q}_2^\ast$.  
From (\ref{eqn:hoge}), we roughly have
\[
\mathsf{Q}_1^\ast\mathbf{\underline{U}}\mathsf{Q}_1
=\mathsf{Q}_1^\ast[\vct{y}_1,\ldots,\vct{y}_{N-1}]\mathsf{Q}_2\mathsf{\Sigma}^{-1}.
\]
This right-hand side can be computed directly from the data and is hence thought of as an approximation of the action of the Koopman operator in the projected space with $\mathsf{Q}_1$, that is, $m$-dimensional space.  
For data with high spatial dimension in fluid applications, namely $m\gg N$, DMD has scored a great success for characterizing complex fluid fields \cite{Schmid_JFM656,Chen_JNLS22,Tu_JCD1}.  
For data with low spatial dimension ($m\ll N$), the approximation might be not enough to gaining the dynamical information (frequencies, modes, etc.) of $\vct{y}_k$. 

\vspace*{2mm}
\noindent{\bf (c)~Vector Prony analysis}

Finally, the so-called \emph{vector Prony analysis} \cite{Susuki_CDC15} is introduced and connected to the Arnoldi-type algorithm.  
Prony analysis is widely used in applications and is a standard technique to reconstruct a scalar function described by a sparse sum of exponentials from finite function values \cite{Hildebrand:1956,Hauer_IEEETPS5,Plonka:2014,Nabavi_ACC14}.  
Since the family of exponentials is utilized, as clearly stated in \cite{Hauer_IEEETPS5}, the Prony analysis is intended to investigating dynamic data as a solution of \emph{linear} differential equation.  
The authors of \cite{Peter_IP29} generalize the Prony analysis for reconstruction of sparse sums of eigenfuntions of linear operators such as the differential operator.  

Here, let us introduce the procedure of vector Prony analysis.  
The current problem is to find the best approximation of the data (\ref{eqn:data_N}) with the $M$ complex values:
\begin{equation}
\vct{y}_k=\sum_{j=1}^{M}\hat{\lambda}^k_j\hat{\vct{V}}_j,\makebox[2em]{}
k=0,\ldots,N-1,
\end{equation}
where we call $\hat{\lambda}_j\in\bbC$ the $j$-th \emph{Prony value} and $\hat{\vct{V}}_j\in\bbC^m$ the associated $j$-th \emph{Prony vector}.  
The classical Prony analysis provides a rigorous way to obtain $M=N$ pairs of the two Prony quantities from $2N$ sampled \emph{scalars}, namely $m=1$.
The authors of \cite{Susuki_CDC15} extended it and provided the following algorithm for obtaining $M=N$ pairs of the Prony values $\hat{\lambda}_j$ and vectors $\hat{\vct{V}}_j$ from $2N$ sampled \emph{vectors} $\{\vct{y}_0,\ldots,\vct{y}_{2N-1}\}$:  
\begin{enumerate}
\item[(i)] Determine the $N$ coefficients $p_k$ of the \emph{Prony polynomial} defined as
\[
p(z):=\prod^{N}_{j=1}(z-\hat{\lambda}_j)=\sum^{N-1}_{k=0}p_kz^k+z^N,\makebox[2em]{}
z\in\bbC.
\]
Clearly, the $N$ roots of the Prony polynomial coincide with the $N$ Prony values.  
Thus, for $\ell=0,\ldots,N-1$, we have
\[
\sum^{N}_{k=0}p_k\vct{y}_{k+\ell}
=\sum^{N}_{k=0}p_k
\left(\sum^{N}_{j=1}\hat{\lambda}^{k+\ell}_j\hat{\vct{V}}_j\right)
=\sum^{N}_{j=1}\hat{\lambda}^\ell_j\hat{\vct{V}}_j
\left(\sum^{N}_{k=0}p_k\hat{\lambda}^k_j\right)
=\sum^{N}_{j=1}\hat{\lambda}^\ell_j\hat{\vct{V}}_j\,p(\hat{\lambda}_j)
=\vct{0}.
\]
This induces
\[
\sum^{N-1}_{k=0}p_k\vct{y}_{k+\ell}=-\vct{y}_{\ell+N},\makebox[2em]{}
\ell=0,\ldots,N-1.
\]
Therefore, to determine the coefficients $p_k$, we derive the following linear equation that can be \emph{overdetermined} because of $m\geq 1$:
\begin{equation}
\mathsf{H}\vct{p}=\vct{b},
\label{eqn:Hp=b}
\end{equation}
where $\mathsf{H}\in\bbR^{(m\cdot N)\times N}$ and we call it \emph{vector Hankel matrix} given by
\[
\mathsf{H}:=\left[
\begin{array}{ccccc}
\vct{y}_0 & \vct{y}_1 & \vct{y}_2 & \cdots & \vct{y}_{N-1}\\
\vct{y}_1 & \vct{y}_2 & \vct{y}_3 & \cdots & \vct{y}_{N}\\
\vct{y}_2 & \vct{y}_3 & \vct{y}_4 & \cdots & \vct{y}_{N+1}\\
\vdots & \vdots & \vdots & \ddots & \vdots\\
\vct{y}_{N-1} & \vct{y}_{N} & \vct{y}_{N+1} & \cdots & \vct{y}_{2N-2}
\end{array}
\right],~
\vct{p}=\left[\begin{array}{c}
p_0\\ p_1\\ p_2\\ \vdots\\ p_{N-1}
\end{array}\right],~
\vct{b}:=-\left[\begin{array}{c}
\vct{y}_N\\ \vct{y}_{N+1}\\ \vct{y}_{N+2} \\ \vdots\\ \vct{y}_{2N-1}
\end{array}\right].
\]
Now, by assuming $\mathsf{H}$ to be of full column rank, the true solution of the target coefficients is explicitly expressed as follows:
\[
\vct{p}=(\mathsf{H}^\top\mathsf{H})^{-1}\mathsf{H}^\top\vct{b}.
\]
If this is not the case, use the Moore-Penrose pseudo-inverse of $\mathsf{H}$. 
\item[(ii)] Find the $N$ roots $\hat{\lambda}_j$ of the Prony polynomial.   
This is equivalent to locating the $N$ eigenvalues of the companion matrix $\mathsf{C}'_N$:
\[
\mathsf{C}'_N:=\left[
\begin{array}{ccccc}
0 & 0 & \cdots & 0 & -p_0\\
1 & 0 & \cdots & 0 & -p_1\\
0 & 1 & \cdots & 0 & -p_2\\
\vdots & \vdots & \ddots & \vdots & \vdots\\
0 & 0 & \cdots & 1 & -p_{N-1}\\
\end{array}
\right].
\]
\item[(iii)] and (iv)~~Same as (iii) and (iv) of the Arnoldi-type algorithm.
\end{enumerate}

Here we provide a geometrical interpretation of vector Prony analysis \cite{Raak_CDC16}.  
By following the argument in the beginning of Section~\ref{subsec:KM-algo}, define the function space $\mathcal{K}'_N$ generated by the operation of \emph{time-shifting} as
\[
\mathcal{K}'_N:=\mathrm{span}\left\{
\left[
\begin{array}{c}
\mathbf{\underline{U}}^k\vct{f} \\ \mathbf{\underline{U}}^{k+1}\vct{f} \\ \vdots \\\noalign{\vskip 1mm}
\mathbf{\underline{U}}^{k+N-1}\vct{f}
\end{array}
\right]; k=0,\ldots,N-1
\right\}.
\]
Then, by letting $\bP'_N:\cF^m\to\mathcal{K}'_N$ be a projection from the original space $\cF^m$ onto $\mathcal{K}'_N$, the following operator
\[
\bP'_N\mathbf{\underline{U}}: \mathcal{K}'_N\to\mathcal{K}'_N
\]
is still finite-dimensional and has a matrix representation with the same dimension of $\mathsf{C}_N$ in the $\{[(\mathbf{\underline{U}}^k\vct{f})^\top,$ $\ldots,(\mathbf{\underline{U}}^{k+N-1}\vct{f})^\top]^\top\}$-basis, which we have denoted by $\mathsf{C}'_N:\bbC^N\to\bbC^N$.  
Needless to say, we have implicitly assumed that the $N$ new functions (``fictive" observables) are linearly independent; this would be more relevant than in $\mathcal{K}_N$ with a single $\mathbf{\underline{U}}^k\vct{f}$.  
Now, denoting the vector Hankel matrix $\mathsf{H}$ by the new Krylov subspace $\mathsf{K}'_{N}$ and introducing $\vct{r}\in\bbR^{m\cdot N}$ as a residual with $\vct{r}\bot\,\mathrm{span}\{\mathsf{K}'_{N}\}$, we rewrite (\ref{eqn:Hp=b}) as follows:
\begin{equation}
\vct{r}=\vct{b}-\mathsf{K}'_N\vct{p}.
\label{eqn:p'}
\end{equation}
In analogy with (\ref{eqn:c}) in Arnoldi-type algorithm, we are able to state that in vector Prony analysis, $\vct{p}$ is chosen to minimize the $\bbC^{m\cdot N}$-norm of $\vct{r}$ in (\ref{eqn:p'}), corresponding to choosing the projection operator $\mathbf{P}'_{N}$ so that $\mathbf{P}'_{N}\mathbf{\underline{U}}^{N}\vct{f}$ is the least-squares approximation to $\mathbf{\underline{U}}^{N}\vct{f}$ at $\vct{x}_0\in\mathbb{X}$ satisfying $\vct{y}_0=\vct{f}(\vct{x}_0)$.   
By comparison with $\mathsf{K}_N\in\bbR^{m\times N}$, the rank-deficiency of $\mathsf{K}'_{N}\in\bbR^{(m\cdot N)\times N}$ when choosing $\vct{p}$ can be avoided even in the case of data with low spatial dimension ($m\ll N$).  
Thus, the unique solution of $\vct{p}$, in other words, an unique decomposition of $\vct{y}_k$ can be derived.  
The possibility of avoiding the rank-deficiency is a result of time-shifting in which we are able to utilize more data for computing $\mathsf{C}'_N$.  
The matrix $\mathsf{C}'_N$ is thought of as an approximation to the action of the Koopman operator on the fictive finite-dimensional space $\mathcal{K}'_N$.  

In summary, we state mathematically that the difference of the Arnoldi-type algorithm and vector Prony analysis for KMD is the choice of finite-dimensional function spaces on which the original Koopman operator is projected.  
Both the derived matrices are thought of as the least-squares approximations of the corresponding action of the Koopman operator.  
Technically, the difference is just the number of snapshots necessary for computation: to obtain $N$ pairs of KMs and KEs, $N+1$ snapshots are needed for Arnoldi-type algorithm, while $2N$ snapshots for vector Prony analysis.  
A comparison of the two algorithms using realistic power system data is presented in \cite{Raak_CDC16}.

\subsubsection{Coherency notion}
\label{subsec:KM-coherency}

We define the notion of \emph{coherency} in the context of KM.  
The case of oscillatory KM, in which the KE has a non-zero imaginary part, is now addressed because the study on coherency identification in power systems normally deals with oscillatory responses following a disturbance.  
For an oscillatory pair of KMs $\vct{V}_{i_1}$ and $\vct{V}_{i_2}$, called the KM pair $\{i_1,i_2\}$, with KEs $\lambda_{i_1}=|\lambda_i|\ee^{\ii 2\pi\nu_{i_1}}$ and $\lambda_{i_2}=\lambda^\ast_{i_1}=|\lambda_{i_1}|\ee^{-\ii 2\pi\nu_{i_1}}$, the associated modal dynamics are defined as
\begin{eqnarray}
\vct{f}^{\{i_1,i_2\}}(\vct{x}[k])
&:=& \lambda^k_{i_1}\phi_{i_1}(\vct{x}[0])\vct{V}_{i_1}+\lambda_{i_2}^k\phi_{i_2}(\vct{x}[0])\vct{V}_{i_2}
\nonumber\\
&=& \lambda^k_{i_1}\phi_{i_1}(\vct{x}[0])\vct{V}_{i_1}+(\lambda^\ast_{i_1})^k\{\phi_{i_1}(\vct{x}[0])\vct{V}_{i_1}\}^\mathrm{c}
\nonumber\\
&=& 2|\lambda_{i_1}|^k
\left[
\begin{array}{c}
A_{i_11}\cos(2\pi k\nu_{i_1}+\alpha_{i_11})\\
\vdots\\
A_{i_1m}\cos(2\pi k\nu_{i_1}+\alpha_{i_1m})
\end{array}
\right],
\label{eqn:te2}
\end{eqnarray}
where we have used the \emph{amplitude coefficient} $A_{ij}$ and \emph{initial phase} $\alpha_{ij}$, defined as
\begin{equation}
A_{ij}:=|\phi_i(\vct{x}[0])V_{ij}|,\makebox[1em]{}
\alpha_{ij}:=\mathrm{Arg}(\phi_i(\vct{x}[0])V_{ij})\makebox[1em]{}
(V_{ij}: \textrm{$j$-th component of $\vct{V}_i$}).
\label{eqn:te}
\end{equation}
Thus, we can say that a set of observables $\mathbb{I}\subseteq\{1,\ldots,m\}$ is \emph{coherent} with respect to the KM pair $\{i_1,i_2\}$ if the amplitude coefficients $A_{i_1j}$ are the same for all $j\in\mathbb{I}$, and the initial phases $\alpha_{i_1j}$ are also the same.  
It is noted that the definition is strict compared with the definitions of slow-coherency \cite{Avramovic_AUTO16,Winkelman_IEEETPAS100} and near-coherency \cite{Sastry_IEEETAC26}, because it does not admit any finite, constant phase difference of oscillations along the system's  dynamics.  
A relaxed notion of the coherency is introduced in \cite{Marko_CHAOS22,Georgescu_EB86,Raak_IEEETPWRS2015} with a small positive parameter $\epsilon$: if for the two observables $f_{j_1}$ and $f_{j_2}$ in $\vct{f}$, the initial phases satisfy $|\alpha_{i_1j_1}-\alpha_{i_1j_2}|<\epsilon$, then the two observables are called $\epsilon$-\emph{phase-coherent} with respect to the KM pair $\{i_1,i_2\}$.  
Below, we will sometime denote the KM pair $\{i_1,i_2\}$ by $i_1$ for simplicity of the notation.

\subsection{Model-order reduction}
\label{subsec:MOR}

By starting at \cite{Mauroy_PD261}, we look at a connection between spectrum of the Koopman operator and canonical equations of motion, and we then provide an idea on model-order reduction of nonlinear systems that is  partly demonstrated in \cite{Susuki_IEEETPWRS27} and Section~\ref{sec:precursor}. 

\subsubsection{Hamilton's canonical form and Koopman eigenfuctions}
\label{subsec:MOR-1}

First of all, let us review the beginning of Hamiltonian mechanics based on \cite{Arnold:1989}.  
The so-called Hamilton's equation of $n$-degree-of-freedom motions is the following: 
for displacement variables $\vct{q}=[q_1,\ldots,q_n]^\top\in\mathbb{M}$ (configuration manifold) and momentum variables $\vct{p}=[p_1,\ldots,p_n]^\top\in\mathrm{T}^\ast_q\mathbb{M}$ (cotangent space of $\mathbb{M}$ at $\vct{q}$),
\begin{equation}
\frac{d}{dt}q_j(t)=\frac{\DD}{\DD p_j}H(\vct{p}(t),\vct{q}(t)),\makebox[2em]{}
\frac{d}{dt}p_j(t)=-\frac{\DD}{\DD q_j}H(\vct{p}(t),\vct{q}(t)),\makebox[2em]{}j=1,\dots,n,
\label{eqn:H}
\end{equation}
where $H:\mathrm{T}^\ast\mathbb{M}~\textrm{(cotangent bundle of $\mathbb{M}$)} \to\bbR$ is the \emph{Hamiltonian} that is normally related to energy of the system.  
If the dynamical system (\ref{eqn:H}) is integrable \cite{Arnold:1989}, then the system is written as the famous form of \emph{action-angle} variable representation: for angle variables $\vct{\theta}=[\theta_1,\ldots,\theta_n]^\top\in\mathbb{T}^n$ and action variables $\vct{I}=[I_1,\ldots,I_n]^\top\in\bbR^n_{\geq 0}$, 
\begin{equation}
\frac{d}{dt}\theta_j(t)=\mathit{\Omega}_j(\vct{I}(t)),\makebox[2em]{}
\frac{d}{dt}I_j(t)=0,\makebox[2em]{}j=1,\dots,n
\label{eqn:H_aa}
\end{equation}
where $\mathit{\Omega}_j: \bbR^n_{\geq 0}\to\bbR$ is the action-dependent function of angular frequency.  
In this, the $n$ action variables $I_1,\ldots,I_n$ are called $n$ constants of motion because their time differentiations are equal to zero.  
The discrete-time version of (\ref{eqn:H_aa}) is given by
\begin{equation}
\theta_j[k+1]=\theta_j[k]+\mathit{\Omega}_j(\vct{I}[k]),\makebox[2em]{}
I_j[k+1]=I_j[k],\makebox[2em]{}j=1,\dots,n.
\label{eqn:H_aa_dds}
\end{equation}
Also, the following perturbed form of (\ref{eqn:H_aa_dds}) has been widely analyzed for modal interaction and instabilities in nearly-integrable Hamiltonian systems: 
\begin{equation}
\left.
\begin{array}{lcl}
\theta_j[k+1] &=& \theta_j[k]+\mathit{\Omega}_j(\vct{I}[k])+F_j(\vct{\theta}[k],\vct{I}[k]),\\\noalign{\vskip 1mm}
I_j[k+1] &=& I_j[k]+G_j(\vct{\theta}[k],\vct{I}[k]),
\end{array}
\right\}
\makebox[2em]{}j=1,\dots,n.
\label{eqn:H_aa_dds_p}
\end{equation}
where $F_j$ and $G_j: \mathbb{T}^n\times\bbR^n_{\geq 0}\to\bbR$ are functions that represent (smooth) perturbations to the integrable motion.  

Here we derive the same system as (\ref{eqn:H_aa_dds}) for the general system (\ref{eqn:dds}) by introducing the Koopman eigenfunctions $\phi_j$ and associated eigenvalues $\lambda_j$.  
First, based on \cite{Mauroy_PD261}, let us consider the Koopman eigenfunctions as a transformation of the state $\vct{x}$ like
\[
z_j=\phi_j(\vct{x}),\makebox[2em]{}z_j\in\bbC,\makebox[2em]{}j=1,2,\ldots
\]
Then, we have
\begin{equation}
z_{j}[k+1]=\phi_j(\vct{x}[k+1])=\phi_j(\vct{T}(\vct{x}[k]))=(\bU\phi_j)(\vct{x}[k])=\lambda_j\phi_j(\vct{x}[k])=\lambda_jz_j[k],
\label{eqn:big-interest_canonical}
\end{equation}
and by taking $\theta_j=\mathrm{Arg}(z_j)$ and $I_j=|z_j|$, we have
\begin{equation}
\theta_j[k+1]=\theta_j[k]+\mathrm{Arg}(\lambda_j),\makebox[2em]{}
I_j[k+1]=|\lambda_j|I_j[k],\makebox[2em]{}j=1,2,\ldots
\label{eqn:big-interest}
\end{equation}
This equation is equivalent to the action-angle representation (\ref{eqn:H_aa_dds}) if $|\lambda_j|=1$ holds.  
The condition of KEs implies that the Koopman operator $\bU$ is unitary, for which the state's dynamics are on an attractor.  
Here we should emphasize that (\ref{eqn:big-interest}) is derived for the general (possibly dissipative) dynamical system (\ref{eqn:dds}) via the Koopman eigenfunctions.  
The vector form of the equation (\ref{eqn:big-interest_canonical}) is derived as follows: for $\vct{z}:=[z_1,z_2,\ldots]^\top$,
\begin{equation}
\vct{z}[k+1]=\mathsf{\Lambda}\vct{z}[k],\makebox[2em]{}
\mathsf{\Lambda}:=\mathrm{diag}(\lambda_1,\lambda_2,\ldots).
\label{eqn:big-interest-2}
\end{equation}
Note that the state space should be taken as $\ell_2$ or $\ell_\infty$ under a certain condition of the infinite-dimensional matrix $\mathsf{\Lambda}$.  

Here, if the {\color{black}stable} dynamics observed have only a finite number of peaks in the frequency domain, they would be dominated by KEs {\color{black}which magnitudes are less than or equal to unity}. 
When we sort the KEs in the \emph{slowest} direction, that is,
\[
\color{black}
|\lambda_{j+1}|\leq|\lambda_{j}|\leq|\lambda_1|\leq 1,\makebox[2em]{}j=2,3,...,
\]
the asymptotic component in the dynamics is dominated by several leading KEs, and the level sets of the associated Koopman eigenfunctions form sub-spaces in the state space, on which the asymptotic component is well captured; this concept is developed in \cite{Mauroy_PD261} as \emph{isostable}.

\subsubsection{Koopman spectrum-preserving model-order reduction}
\label{subsec:MOR-2}

The finding of canonical-like form of equations leads to an idea of model-order reduction of the discrete-time system (\ref{eqn:dds}).  
Now, let us suppose that the first $M$ variables $z_1,\ldots,z_M$ in (\ref{eqn:big-interest-2}) are sufficient for capturing the state's dynamics of interest.  
For this, let us introduce a finite-dimensional truncation of the first $M$ Koopman eigenvalues and eigenfunctions represented by a infinite-dimensional matrix with a finite $M$ rank,
\[
\mathsf{Q}_M:=\mathrm{diag}(\underbrace{1,1,\ldots,1}_{M},0,0,\ldots).
\]  
Thus, we have {\color{black}the following system to represent the state's dynamics:
\[
\mathsf{Q}_M\vct{z}[k+1]=\mathsf{\Lambda}\vct{z}[k].
\]
Here, in order to focus on the time-varying part above, by defining $\vct{z}^M:=[z_1,\ldots,z_M]^\top\in\bbC^M$ and $\mathsf{\Lambda}_M:=\mathrm{diag}(\lambda_1,\ldots,\lambda_M)\in\bbC^{M\times M}$, we have
\begin{equation}
\vct{z}^M[k+1]=\mathsf{\Lambda}_M\vct{z}^M[k].
\label{eqn:rom_dds}
\end{equation}
}For the $n$-dimensional state's dynamics in (\ref{eqn:dds}), if $M\ll n$ holds, then (\ref{eqn:rom_dds}) is a relevant reduced-order representation of the original system (\ref{eqn:dds}).  
Since the dominant part of the discrete spectrum is preserved in the procedure, we call it the \emph{Koopman spectrum-preserving} model-order reduction.

The derivation of the reduced-order model (\ref{eqn:rom_dds}) from (\ref{eqn:dds}) is possible by conducting both the KMD for finite data on the state's dynamics and the Petrov-Galerkin projection method \cite{Holmes:1996} applied to the original system (\ref{eqn:dds}).  
The following idea does not require explicit formula of the Koopman eigenfunctions $\phi_j$.  
Now, consider $N$ pairs of KEs and KMs computed from finite-time data on the state's dynamics $\{\vct{y}_k=\vct{x}_k; k=0,1,\ldots\}$ as
\begin{equation}
\vct{x}_k=\sum_{j=1}^{N}\tilde{\lambda}^k_j\vct{\tilde{V}}_j,\makebox[2em]{}
\vct{\tilde{V}}_j:=\phi_j(\vct{x}_0)\vct{V}_j,
\label{eqn:truncation}
\end{equation}
where $k=0,1,\ldots,N-1$.  
Suppose that we sort the $N$ pairs of KEs and KMs such that the first $M(\leq N)$ pairs are sufficient for capturing the finite-time dynamics of interest.  
Then, the following transformation of the state is defined:
\[
\vct{x}=\mathsf{\tilde{V}}\vct{\tilde{z}}^M,\makebox[2em]{}
\vct{\tilde{z}}^M\in\bbC^M,\makebox[2em]{}
\mathsf{\tilde{V}}:=[\vct{\tilde{V}}_1,\ldots,\vct{\tilde{V}}_M]\in\bbC^{n\times M}.
\]
Thus, by substituting the above to the original system (\ref{eqn:dds}), we have
\[
\mathsf{\tilde{V}}\vct{\tilde{z}}^M[k+1]=\vct{T}(\mathsf{\tilde{V}}\vct{\tilde{z}}^M[k+1]),
\]
and by assuming $\mathsf{\tilde{V}}^\ast\mathsf{\tilde{V}}\in\bbC^{M\times M}$ to be regular and using the standard argument of Petrov-Galerkin projection, we have
\begin{eqnarray}
\vct{\tilde{z}}^M[k+1] &=& (\mathsf{\tilde{V}}^\ast\mathsf{\tilde{V}})^{-1}\mathsf{\tilde{V}}^\ast\vct{T}(\mathsf{\tilde{V}}\vct{\tilde{z}}^M[k+1])
\nonumber\\
&=:& \vct{\tilde{T}}(\vct{\tilde{z}}^M[k]).
\label{eqn:rom_dds_PG}
\end{eqnarray}
Here we represent the $j$-th component of $\vct{\tilde{z}}^M$ as $\tilde{I}_j\ee^{\ii\tilde{\theta}_j}$.  
Along the state's dynamics used for the computation of KMs, without the truncation after the first $M$ KMs in (\ref{eqn:truncation}), the variable $\tilde{I}_j$ does not change as time goes on, and $\tilde{\theta}_j$ rotates with a constant speed $\mathrm{Arg}(\tilde{\lambda}_j)$ given by the KE $\tilde{\lambda}_j$.  
In this way, we can regard the right-hand side of (\ref{eqn:rom_dds_PG}) as another formulation of the reduced-order model (\ref{eqn:rom_dds}) that is computed by a combination of the model (\ref{eqn:dds}) and finite-time data $\{\vct{x}_k\}$.  
Furthermore, we can re-write (\ref{eqn:rom_dds_PG}) as follows:
\[
\tilde{I}_j[k+1]\ee^{\ii\tilde{\theta}_j[k+1]}=\tilde{T}_j(\vct{\tilde{I}}[k],\vct{\tilde{\theta}}[k]),\makebox[2em]{}
j=1,\ldots,M,
\]
where $\vct{\tilde{\theta}}=[\tilde{\theta}_1,\ldots,\tilde{\theta}_M]^\top$ and $\vct{\tilde{I}}=[\tilde{I}_1,\ldots,\tilde{I}_M]^\top$, and we have
\begin{equation}
\left.
\begin{array}{ccccc}
\tilde{\theta}_j[k+1] &=& \mathrm{Arg}(\tilde{T}_j(\vct{\tilde{I}}[k],\vct{\tilde{\theta}}[k])) 
&=:& \tilde{F}_j(\vct{\tilde{I}}[k],\vct{\tilde{\theta}}[k])),
\\\noalign{\vskip 1mm}
\tilde{I}_j[k+1] &=& \sqrt{\tilde{T}^\ast_j(\vct{I}[k],\vct{\theta}[k])\tilde{T}_j(\vct{I}[k],\vct{\theta}[k])}
&=:& \tilde{G}_j(\vct{\tilde{I}}[k],\vct{\tilde{\theta}}[k])),
\end{array}
\right\}
\label{eqn:rom_aa_dds_PG}
\end{equation}
or simply
\begin{equation}
\left.
\begin{array}{ccccl}
\vct{\tilde{\theta}}[k+1] &=& \vct{\tilde{F}}(\vct{\tilde{I}}[k],\vct{\tilde{\theta}}[k])) 
&:=& [\tilde{F}_1(\vct{\theta},\vct{I}),\ldots,\tilde{F}_M(\vct{\theta},\vct{I})]^\top,
\\\noalign{\vskip 1mm}
\vct{\tilde{I}}[k+1] &=& \vct{\tilde{G}}(\vct{\tilde{I}}[k],\vct{\tilde{\theta}}[k])) 
&:=& [\tilde{G}_1(\vct{\theta},\vct{I}),\ldots,\tilde{G}_M(\vct{\theta},\vct{I})]^\top.
\end{array}
\right\}
\label{eqn:rom_aa_dds_PG2}
\end{equation}
The right-hand sides of (\ref{eqn:rom_aa_dds_PG}) are regarded as terms in an action-angle formulation of the reduced-order model (\ref{eqn:rom_dds}) that are analogue to (\ref{eqn:H_aa_dds_p}).

\section{Coherency identification of coupled swing dynamics}
\label{sec:coherency}

This section provides the first application of KMD to power systems technology---identification of coherency of swing dynamics in coupled generator networks \cite{Susuki_IEEETPWRS26}.   
The so-called \emph{coherency identification} is related to short-term swing or transient stability as an ability of a power system to maintain synchronism when subjected to a large disturbance.  
The identification is to find a group of synchronous generators swinging together with the in-phase motion and is a basis of development of reduced-order models, which is traditionally used to reduce computational effort and currently employ on-line stability assessment, and dynamical system analysis of loss of transient stability (see \cite{Susuki_JNLS09}).  
Many groups of researchers have developed methods for coherency identification:  see \cite{Podmore_IEEETPAS97,Avramovic_AUTO16,Winkelman_IEEETPAS100,Yusof_IEEETPS8,Sastry_IEEETAC26,Anaparthi_IEEETPS20} and references of \cite{Susuki_IEEETPWRS26}.  
In this section, we apply the technique in Sections~\ref{subsec:KM-algo}(a) and \ref{subsec:KM-coherency} to data on short-term swing dynamics in the New England 39-bus test system (NE system) \cite{Pai:1989} and demonstrate how the technique works for identification of modes and coherency.

\subsection{New England test system and coupled swing dynamics}
\label{subsec:coherency-NE}

\begin{figure*}[t] 
\centering
\subfigure[One-line diagram of the NE system]{%
\includegraphics[width=0.47\textwidth]{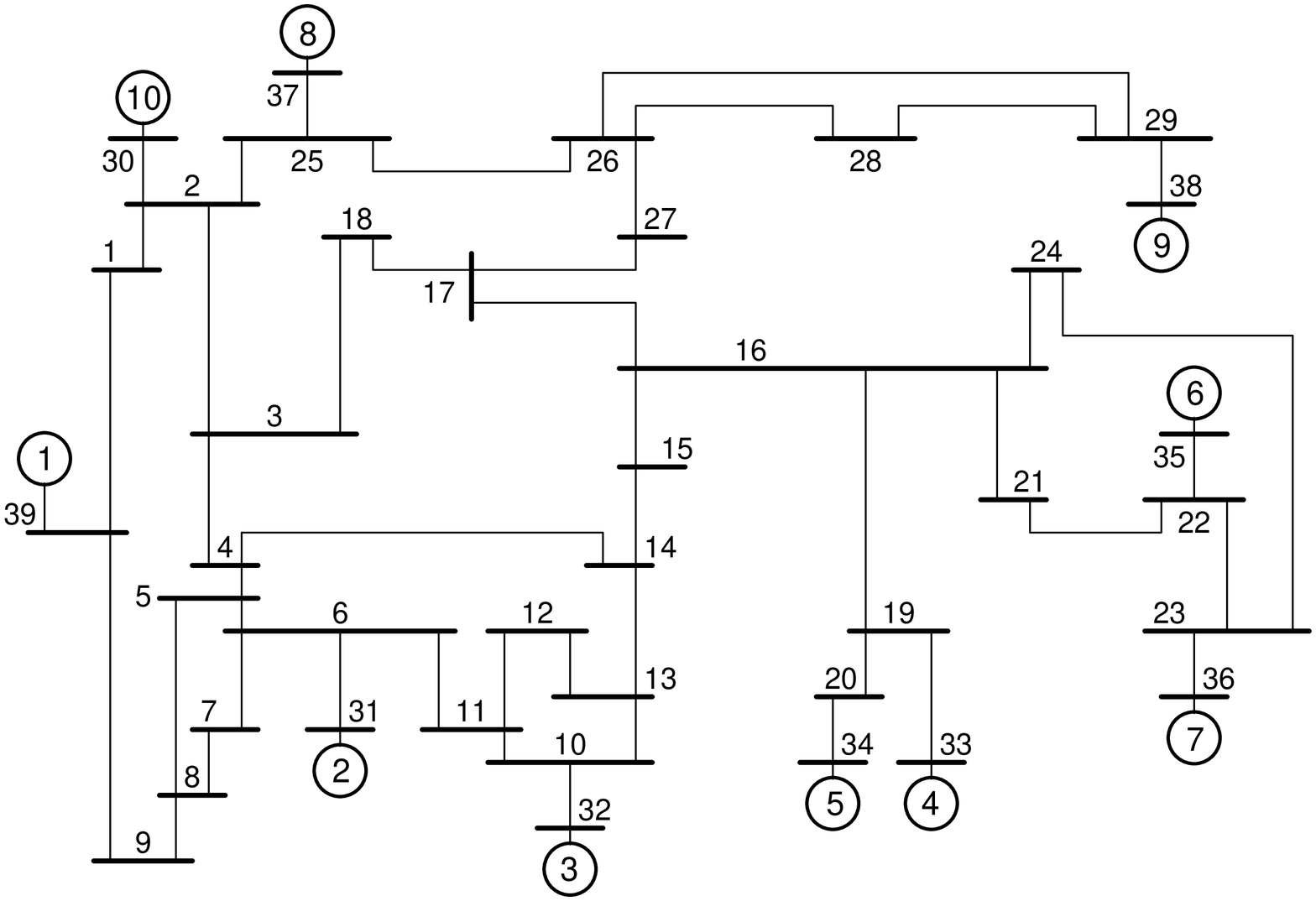}
\label{fig:NEsystem}
}%
\hspace*{4mm}
\subfigure[Time responses of generators 2--10 as the solution of (\ref{eqn:classical_NEsystem}) for the initial condition (\ref{eqn:ic})]{%
\includegraphics[width=0.45\textwidth]{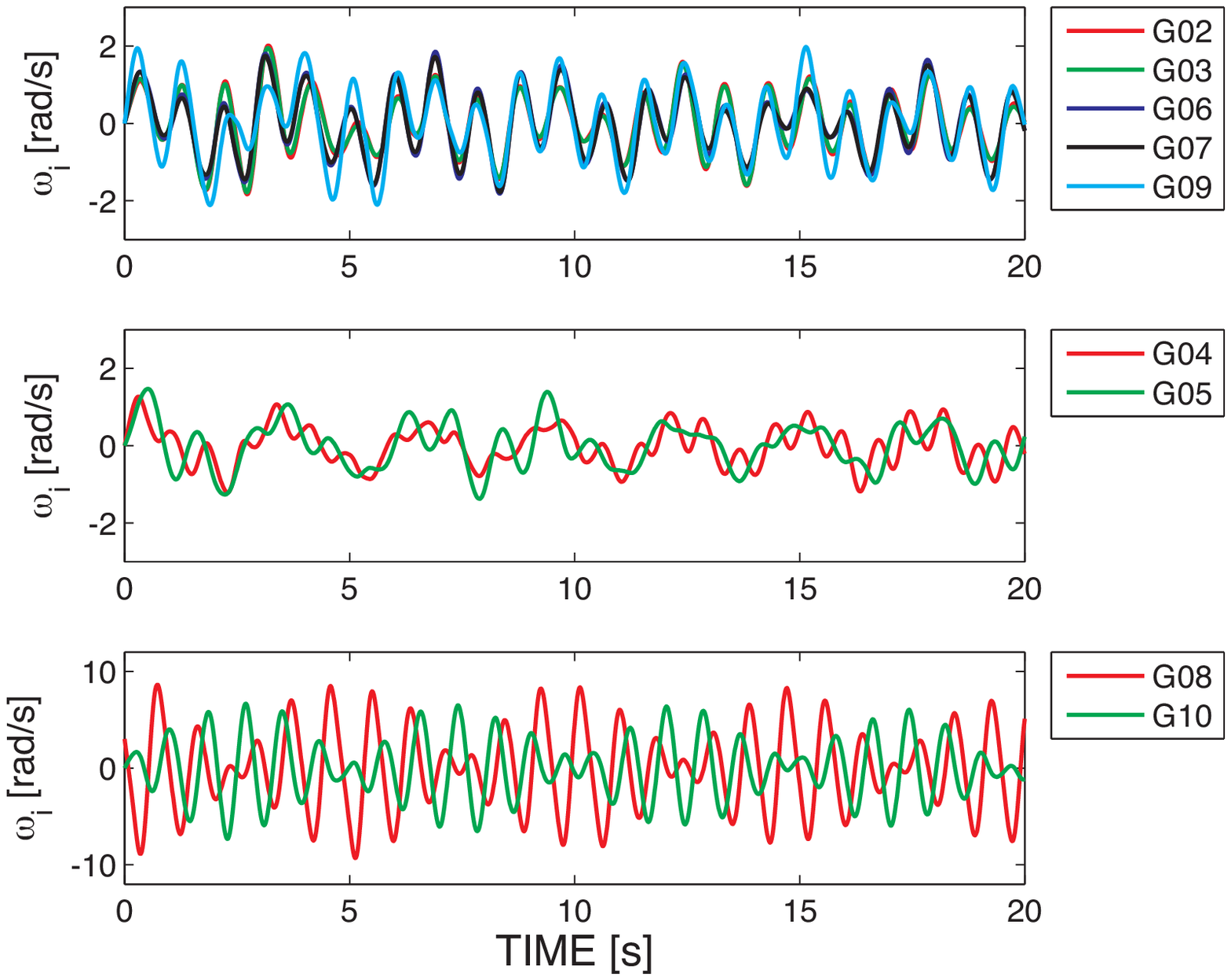}
\label{fig:waveform1}
}%
\caption{%
Coupled swing dynamics in the New England 39-bus test system (NE system) 
\copyright 2011 IEEE
}
\end{figure*}

The NE system is a well-known benchmark model for power systems research shown in Fig.\,\ref{fig:NEsystem} and contains 10 generation units (equivalently 10 synchronous generators, circled numbers in the figure), 39 buses, and AC transmission lines.  
Most of the buses have constant active and reactive power loads.   
See \cite{Pai:1989} for the details of the system.  

Here, we introduce the equations of motion of generators in the NE system.  
Assume that bus 39 is the infinite bus\footnote{A source of voltage constant in phase, magnitude, and frequency, and not affected by the amount of current withdrawn from it}.  
The short-term swing dynamics of generators 2--10 are represented by the following nonlinear differential equations \cite{Machowski:1997}: 
\begin{equation}
\makebox[-1.0em]{}\left.
\begin{array}{rcl}
{\displaystyle \frac{d\delta_i}{dt}\,}
&=& 
\omega_i,
\\\noalign{\vskip +1mm}
{\displaystyle \frac{H_i}{\pi f\sub{b}}\frac{d\omega_i}{dt}} 
&=& 
{\displaystyle -D_i\omega_i+P_{\mathrm{m}i}-G_{ii}E^2_i
-\!\!\sum_{j=1,j\neq i}^{10}\!\!
E_iE_j\left\{G_{ij}\cos(\delta_i-\delta_j)+B_{ij}\sin(\delta_i-\delta_j)\right\},
}
\end{array}
\right\}
\label{eqn:classical_NEsystem}
\end{equation}
where the integer label $i=2,\ldots,10$ stands for generator $i$.  
The variable $\delta_i$ is the angular position of rotor in generator $i$ with respect to bus 1 and is in radians [rad].  
The variable $\omega_i$ is the deviation of rotor speed in generator $i$ relative to that of bus 1 and is in radians per second [rad/s].  
We set the variable $\delta_1$ to a constant, because bus 39 is assumed to be the infinite bus.  
The parameters $f\sub{b}$, $H_i$, $D_i$, $P_{\mathrm{m}i}$, $E_i$, $G_{ii}$, $G_{ij}$, and $B_{ij}$ are in per unit system except for $H_i$ and $D_i$ in seconds [s], and for $f\sub{b}$ in Hertz [Hz].  
The mechanical input power $P_{\mathrm{m}i}$ to generator $i$ and its internal voltage $E_i$ are normally constant in the short-term regime \cite{Machowski:1997}.  
The parameter $H_i$ is the per unit time inertia constant of generator $i$, and $D_i$ its damping coefficient.  
The parameter $G_{ii}$ is the internal conductance, and $G_{ij}+\ii B_{ij}$ is the transfer impedance between generators $i$ and $j$.  
Electrical loads are modeled as constant impedances.  

Now, we present an example of data on short-term swing dynamics in the NE system.  
Fig.\,\ref{fig:waveform1} shows the time responses of rotor speed deviations $\omega_i$ under the initial condition:
\begin{equation}
(\delta_i(0),\omega_i(0))=
\left\{
\begin{array}{ll}
(\delta^\ast_i+1.5\U{rad},3\U{\,rad/s}) & i=8,\\
(\delta^\ast_i,0\U{rad/s}) & \mathrm{else,}
\end{array}
\right.
\label{eqn:ic}
\end{equation}
where $\delta_i^\ast$ represents the value of stable equilibrium point of (\ref{eqn:classical_NEsystem}). 
The initial condition physically corresponds to a local disturbance at generator 8.  
The details of the numerical setting are in \cite{Susuki_IEEETPWRS26}.  
The generators do not show any divergence motion in the figure, that is, they do not show any loss of transient stability for the selected disturbance.  
Generators 2, 3, 6, and 7 show coherent swings excited by the local disturbance.  
We call these generators the \emph{coherent} group.  
The other generators show incoherent swings in the figure.  
Generator 9 shows swings similar in frequency and phase to the coherent group, but the amplitude of swings is a little larger.  
Generators 8 and 10 have swings of larger amplitudes than the others, because the initial condition is localized at generator 8, and the two generators are electrically close.

\subsection{Coherency identification via Koopman mode decomposition}
\label{subsec:coherency-KMD}

We compute the KMs and KEs for the data on coupled swing dynamics shown in Fig.\,\ref{fig:waveform1}.  
The computation is done with the two different algorithms: Fourier-based formula (\ref{eqn:projection}) and Arnoldi-type algorithm in Section~\ref{subsec:KM-algo}(a).  
For computation we need to choose the observable $\vct{f}(\vct{\delta},\vct{\omega})$ where $\vct{\delta}=(\delta_2,\ldots,\delta_{10})^\top$ and $\vct{\omega}=(\omega_2,\ldots,\omega_{10})^\top$.  
In this section, we use the variables of rotor speed deviations, $\vct{\omega}$, as the observable:  $\vct{f}(\vct{\delta},\vct{\omega})=\vct{\omega}$.  
This observable has a clear physical meaning in power systems:  one measures rotor speeds or frequencies for every generation plant.  
We use the simulation output shown in Fig.\,\ref{fig:waveform1} that extracts $\{\vct{\omega}(nT)\}_{n=0}^{N}$, where the uniform sampling period $T=1/(50\U{Hz})$ and the number of snapshots $N+1=1001$. 

\begin{figure}[t] 
\begin{center}
\includegraphics[width=0.45\textwidth]{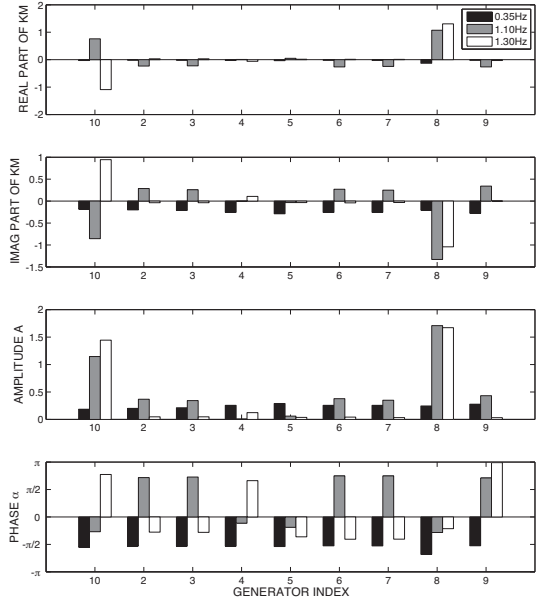}
\caption{%
Numerical results of the terms $\phi_i(\vct{x}[0])\vct{V}_i$ using the Fourier-based formula (\ref{eqn:projection}) under $\nu=(0.35\U{Hz})T$, $(1.10\U{Hz})T$, and $(1.30\U{Hz})T$.  
The amplitude coefficients $A_{ji}$ and initial phases $\alpha_{ji}$, defined in (\ref{eqn:te}), are shown. 
Reproduced from \cite{Susuki_IEEETPWRS26}
}%
\label{fig:KM5}
\end{center}
\end{figure}

A Fourier analysis of the waveforms in Fig.\,\ref{fig:waveform1} shows that generators 2, 3, 6, and 7 have the similar shape of spectrum with peak frequencies 0.35\,Hz and 1.10\,Hz, and that generators 8 and 10 have the peak frequencies 1.10\,Hz and 1.30\,Hz.  
Hence we compute the terms including KMs, $\phi_i(\vct{x}[0])\vct{V}_i$, using the projection operator $\bP^\nu$ with $\nu=(0.35\U{Hz})T$, $(1.10\U{Hz})T$, and $(1.30\U{Hz})T$.  
We use the finite-time approximation of (\ref{eqn:projection}) from $k=0$ to $N$.  
The numerical results are shown in Fig.\,\ref{fig:KM5}.  
The amplitude coefficients and initial phases, which are defined in (\ref{eqn:te}), are also shown.  
For 0.35\,Hz, the values of amplitude coefficients are close for each of the generators, and their initial phases are also close except for generator 8.  
The generators except for 8 hence show in-phase swings with 0.35\,Hz.  
For 1.10\,Hz, the values of amplitude coefficients and initial phases are close for generators 2, 3, 6, 7, and 9, and hence they show in-phase swings with 1.10\,Hz.  
The two KMs with 0.35\,Hz and 1.10\,Hz capture the coherent motion of generators 2, 3, 6, 7, and 9.  For 1.10\,Hz and 1.30\,Hz, the amplitude coefficients for generators 8 and 10 are larger than the others.  
These two KMs capture the large swings of generators 8 and 10 observed in Fig.\,\ref{fig:waveform1}.  
Thus, we can extract spatial modes oscillating with a single frequency directly from the data on  simulation output of the NE system.  

\begin{figure*}[t] 
\centering
\subfigure[Empirical Ritz values $\tilde{\lambda}_j$. 
The color varies smoothly from red to white, depending on the norm of the corresponding mode \copyright 2011 IEEE]{%
\includegraphics[width=0.35\textwidth]{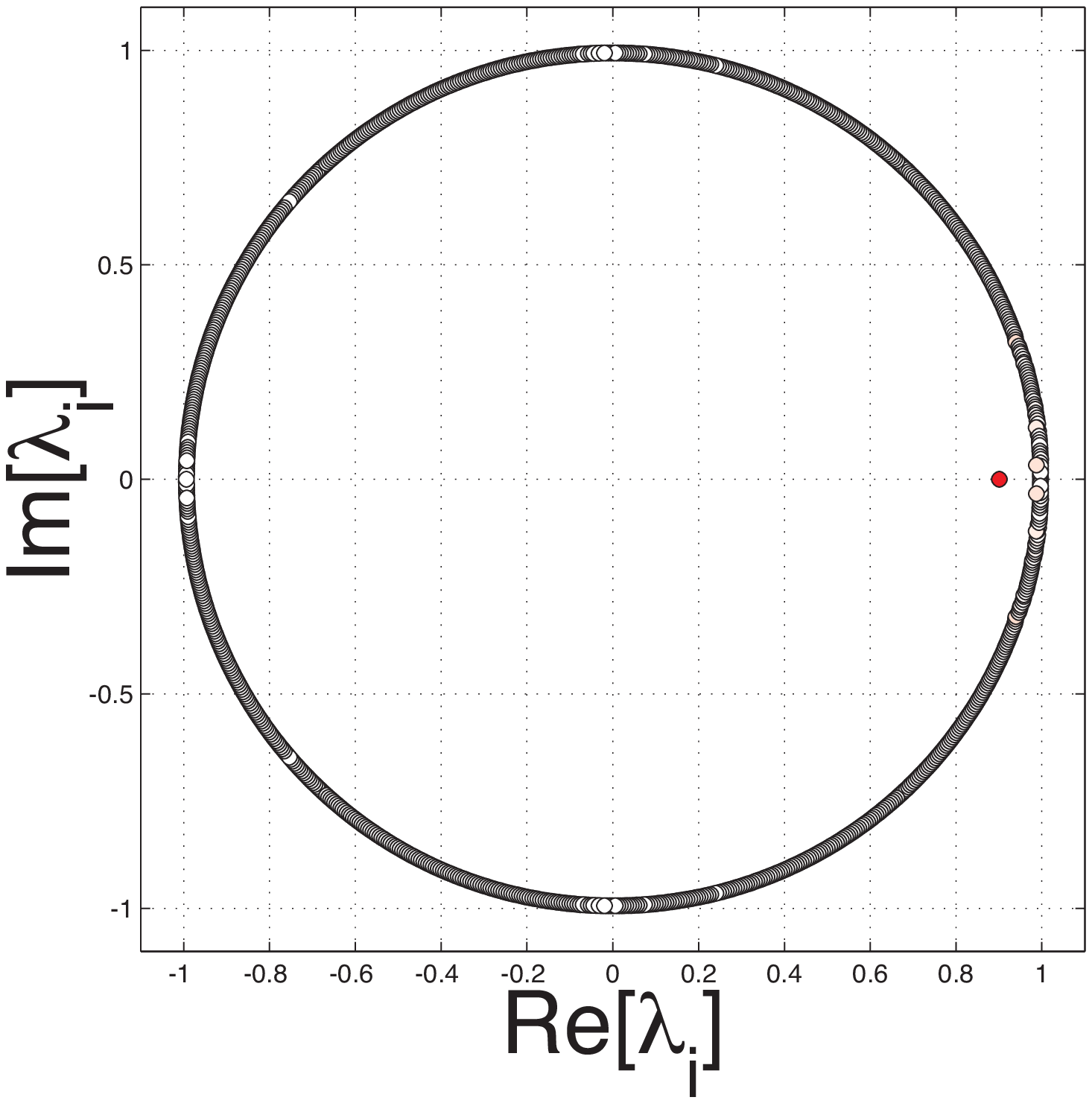}
\label{fig:KM1}}%
\hspace*{12mm}
\subfigure[Empirical Ritz vectors $\tilde{\vct{V}}_j$ ($j=7,8,9$) in Tab.\,\ref{tab:KM}.  
The amplitude coefficients $A_{ji}$ ($i=2,\ldots,10$) and initial phases $\alpha_{ji}$, defined in (\ref{eqn:te}), are shown.  Reproduced from \cite{Susuki_IEEETPWRS26}]{%
\includegraphics[width=0.45\textwidth]{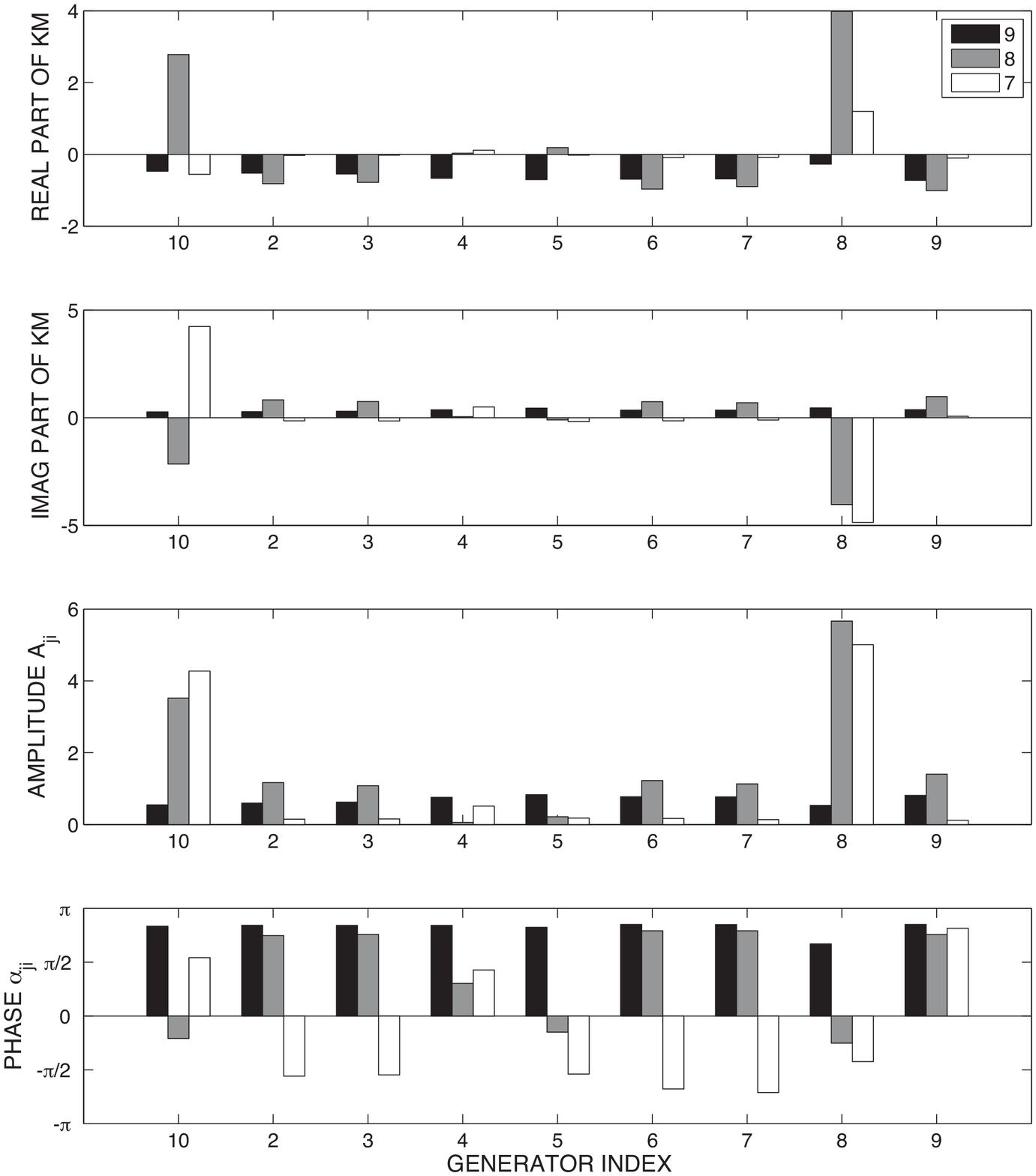}
\label{fig:KM3}}%
\caption{%
Koopman mode decomposition of coupled swing dynamics in the New England test system.  
They are conducted with Arnoldi-type algorithm.}%
\end{figure*}

\begin{table}[t] 
\begin{center}
\caption{%
Numerical results on Koopman eigenvalues and modes obtained with Arnoldi-type algorithm.  
Reproduced from \cite{Susuki_IEEETPWRS26}
}%
\label{tab:KM}
\begin{tabular}{ccccc}\hline
KM pair & Growth Rate & Argument [rad] & Frequency [Hz] & Norm \\
$j$ & $|\tilde{\lambda}_j|$ & & & $||\vct{\tilde{V}}_j||$ \\\hline
1 & 0.9986 & $\pm$0.1701 & 1.3533 & 3.0021\\
2 & 0.9986 & $\pm$0.1438 & 1.1447 & 2.3930\\
3 & 0.9985 & $\pm$0.1009 & 0.8028 & 0.7039\\
4 & 0.9985 & $\pm$0.1300 & 1.0343 & 0.9753\\
5 & 0.9984 & $\pm$0.0931 & 0.7405 & 0.4507\\
6 & 0.9984 & $\pm$0.1130 & 0.8990 & 0.8162\\
{\bf 7} & 0.9983 & $\pm$0.1643 & 1.3078 & 6.6147\\
{\bf 8} & 0.9983 & $\pm$0.1378 & 1.0962 & 7.1941\\
{\bf 9} & 0.9983 & $\pm$0.0468 & 0.3727 & 2.1006\\
10 & 0.9982 & $\pm$0.1836 & 1.4612 & 1.2238\\\hline
\end{tabular}
\end{center}
\end{table}    

\begin{figure}[t]
\centering
\includegraphics[width=0.5\textwidth]{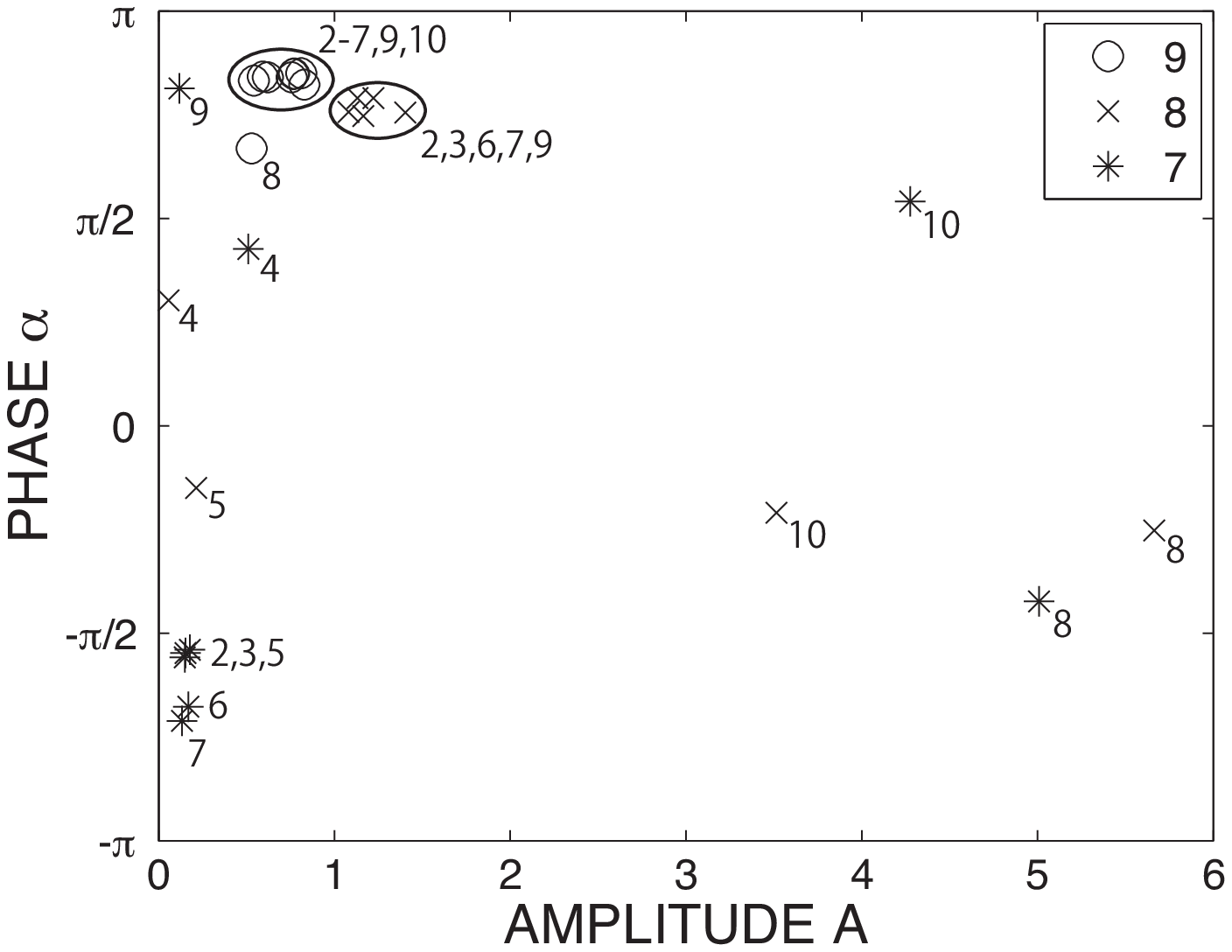}
\caption{%
Distribution of amplitude coefficients $A_{ji}$ and initial phases $\alpha_{ji}$ for the Koopman modes ($j=7,8,9$) shown in Fig.\,\ref{fig:KM3}.  
The numbers inside the figure (e.g. 8,10) denote the integer labels of generators. 
\copyright 2011 IEEE
}%
\label{fig:A}
\end{figure}

Next, we compute the KEs and KMs (empirical Ritz values $\tilde{\lambda}_j$ and associated vectors $\tilde{\vct{V}}_j$) using the Arnoldi-type algorithm.  
Fig.\,\ref{fig:KM1} shows the empirical Ritz values $\tilde{\lambda}_j$.  
The norm of $\vct{r}$ in Step (i) was of order $10^{-12}$.  
Many KMs are obtained and are close to the unit circle $|\tilde{\lambda}_j|=1$.  
Now let us focus on KMs that have both large growth rates $|\tilde{\lambda}_j|$ and large norms of $\tilde{\vct{V}}_j$.  
Such modes represent sustained swing components for the time duration of data and have dominant magnitudes in the time series.  
Tab.\,\ref{tab:KM} shows numerical results on KEs and KMs, which we call KM pair 1 to 10.  
The order of KM pairs in Tab.\,\ref{tab:KM} is based on the magnitudes of growth rates.  
We now pick up KM pair~7 to 9 that have large norms in the table.  
KM pair~1 and 2 have large norms, too.  
But, their frequencies are close to KM pair~7 and 8, respectively.  
Fig.\,\ref{fig:KM3} shows the KMs $\tilde{\vct{V}}_j$ for Mode\,$j$ ($j=7,8,9$).  
The amplitude coefficients $A_{ji}$ ($j=7,8,9, i=2,\ldots,10$) and initial phases $\alpha_{ji}$ are also shown.  
We display the results in order to make it easy to compare this with the Fourier-based results in Fig.\,\ref{fig:KM5}.  
For example, the frequency 1.3078\,Hz of KM pair~7 is close to one of the dominant frequencies for generators 8 and 10, that is, 1.30\,Hz.  
In fact, since the values of amplitude coefficients $A_{7,i}$ in Fig.\,\ref{fig:KM3} are large for generators 8 and 10, the corresponding modal dynamics are localized at these generators.  
Thus, we can decompose the coupled swing dynamics in the NE system into a set of KMs, namely, spatial modes of oscillation with single frequency.  

The decomposition into KMs makes it possible to extract coherent generators in the coupled swing dynamics.  
In fact, the two KM pairs\,8 and 9 capture a coherent motion related to the coherent group of generators.  
These frequencies, 1.0962\,Hz and 0.3727\,Hz, are close to the frequencies of the coherent group, 1.10\,Hz and 0.35\,Hz.  
For KM pair~8, the values of $A_{8,i}$ are close for each of generators 2, 3, 6, 7, and 9, and their initial phases $\alpha_{8,i}$ are also close.  
The distribution of $(A_{8,i},\alpha_{8,i})$ is plotted in Fig.\,\ref{fig:A}.  
The points ($\times$) for generators 2, 3, 6, 7, and 9 are clustered around the coordinate $(1.2,3\pi/4)$.  
Generators 2, 3, 6, 7, and 9 hence show in-phase swings with 1.0962\,Hz.  
For KM pair~9, the values of $A_{9,i}$ and $\alpha_{9,i}$ are close for each of the generators except for generator 8, and hence they show in-phase swings with 0.3727\,Hz.  
The corresponding plots ($\circ$) in Fig.\,\ref{fig:A} are clustered except for generator 8.  
The two KMs capture the coherent motion of generators 2, 3, 6, 7, and 9.  
In this way, we can identify the coherent group observed in Fig.\,\ref{fig:waveform1} by using the decomposition into KMs.   
The plot as Fig.\,\ref{fig:A} provides a systematic way to identify coherent swings and generators by using an automatic clustering algorithm.

In this section, we used the Fourier-based formula (\ref{eqn:projection}) and Arnoldi-type algorithm in Section~\ref{subsec:KM-algo}(a) for cross-check of the computation of KMs and KEs.  
The results for amplitude coefficients and initial phases between Fig.\,\ref{fig:KM5} and Fig.\,\ref{fig:KM3} with KM pair~7 to 9 are compared.  
The results for amplitude coefficients are qualitatively similar.  
However, the results for initial phases are somewhat different, especially at generator 8.  
Also, for the amplitude coefficients, there are quantitative differences between the results in Figs.\,\ref{fig:KM5} and \ref{fig:KM3}.  
These differences might be due to the fact that the Fourier-based formula (\ref{eqn:projection}) assumes the unitary case of the Koopman operator, that is, the dynamics are on an attractor.  
Indeed, the current analysis is performed for off-attractor dynamics.  
However, identification of coherent swings and generators is possible using both the algorithms.  
In fact, the coherent swings of generators 2, 3, 6, 7, and 9 are captured well.

\section{Precursor diagnostic of coupled swing instabilities}
\label{sec:precursor}

This section studies a precursor to phenomena of loss of stability in the coupled swing dynamics \cite{Susuki_IEEETPWRS27}.  
Precursor-based monitoring of instabilities has attracted a lot of interest in terms of prediction:  see \cite{Tamura_IEEETPWRS2,Dobson_IEEETCASI48,Wiesenfeld_JSP38} and references of \cite{Susuki_IEEETPWRS27}.  
The precursor in this section is based on the discovery of \cite{Susuki_JNLS09}, an emergent transmission path of energy from many oscillatory modes to one oscillatory mode that represents an instability phenomenon of interest.
The pathway from high frequency modes to the lowest frequency mode is called the Coherent Swing Instability (CSI).    
The modes can be extracted from sensor data or data provided by simulation outputs of power system oscillations via KMD.  
The CSI transmission path is identified by computation of the so-called action transfer operator that is derived by refining the mathematical model (\ref{eqn:classical_NEsystem}) using the technique in Section~\ref{subsec:MOR}.  
This provides a new technique for monitoring the instability by a combination of practical data, mathematical modeling, and computation.

\subsection{Derivation of action transfer operator} 
\label{sec:3-C}

The authors of \cite{Eisenhower_CDC07} gave an insight into the transmission paths of energy in a nearly-integrable Hamiltonian system by linearizing a perturbed action-angle representation. 
The linearized system provides the time-dependent operator as the system evolves.  
In this section, we will term this operator the \emph{action transfer operator} that can quantify the change of action variables.  
Now, we apply the approach developed in \cite{Eisenhower_CDC07} to the discrete-time system (\ref{eqn:rom_aa_dds_PG2}) and show that by means of KM, the action transfer operator is defined for arbitrary dynamical systems.  
The linearizd system of (\ref{eqn:rom_aa_dds_PG2}) around a state $(\vct{\tilde{\theta}}^\ast,\vct{\tilde{I}}^\ast)$ is represented as
\begin{equation}
\left[
\begin{array}{c}
\vct{\tilde{\theta}}[k+1]\\
\vct{\tilde{I}}[k+1]
\end{array}
\right]
=
\left.
\left[
\begin{array}{cc}
\mathrm{D}_{\tilde{\theta}}\vct{\tilde{F}} & \mathrm{D}_{\tilde{I}}\vct{\tilde{F}}\\
\mathrm{D}_{\tilde{\theta}}\vct{\tilde{G}} & \mathrm{D}_{\tilde{I}}\vct{\tilde{G}}
\end{array}
\right]
\right\vert_{(\vct{\tilde{\theta}}^\ast,\,\vct{\tilde{I}}^\ast)}
\left[
\begin{array}{cc}
\vct{\tilde{\theta}}[k]\\
\vct{\tilde{I}}[k]
\end{array}
\right].
\nonumber
\end{equation}
Since the action variables are often associated with amount of energy contained in every KM, the sub-matrix $\mathrm{D}_{\tilde{I}}\vct{\tilde{G}}(\vct{\tilde{\theta}}^\ast,\vct{\tilde{I}}^\ast)$ provides a quantitative index of the model interaction via energy.  
Thus, we can define the Jacobian matrix $\mathsf{J}(\vct{x}[k])$ that describes the infinitesimal change of $\vct{\tilde{I}}[k]$ at a state $\vct{x}[k]$:
\begin{equation}
{\sf J}(\vct{x}[k]):=\mathrm{D}_{\tilde{I}}\vct{\tilde{G}}(\vct{x}[k])-\mathrm{diag}(1,1,\ldots,1).
\label{eqn:J}
\end{equation}
The right-hand side of (\ref{eqn:J}) can be computed numerically: see Appendix of \cite{Susuki_IEEETPWRS27} in detail.  
The Jacobian matrix is a time-dependent operator and quantifies the change of action variables for the $M$ KMs.  
We term the Jacobian matrix the action transfer operator for the $M$ KMs.  
In this section, by using a concrete example of power system analysis, we will illustrate that the action transfer operator works for identification of a precursor to the instability.

\subsection{Coherent swing instability of New England test system}
\label{subsec:precursor-CSI}

\begin{figure}[t] 
\centering
\includegraphics[width=0.55\textwidth]{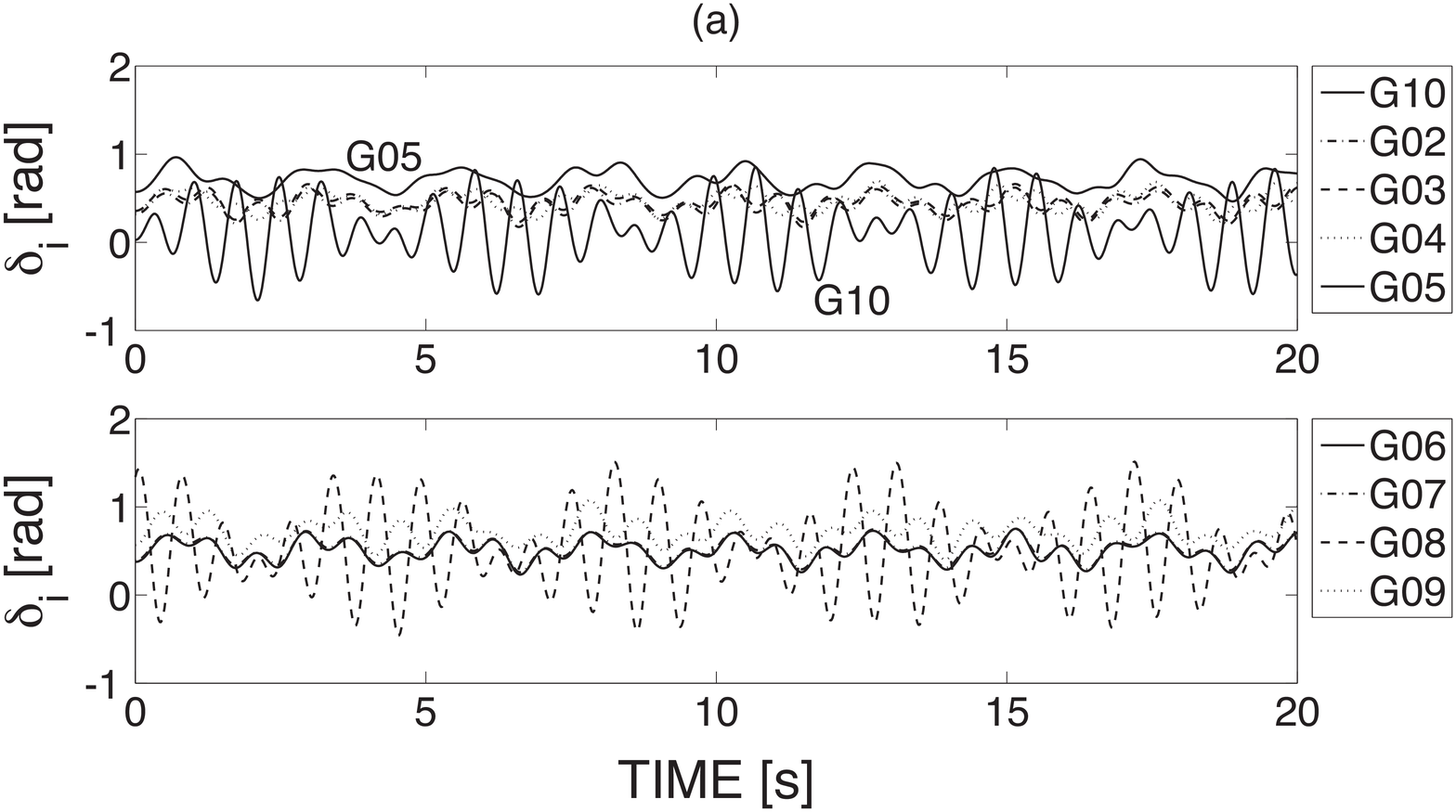}
\includegraphics[width=0.55\textwidth]{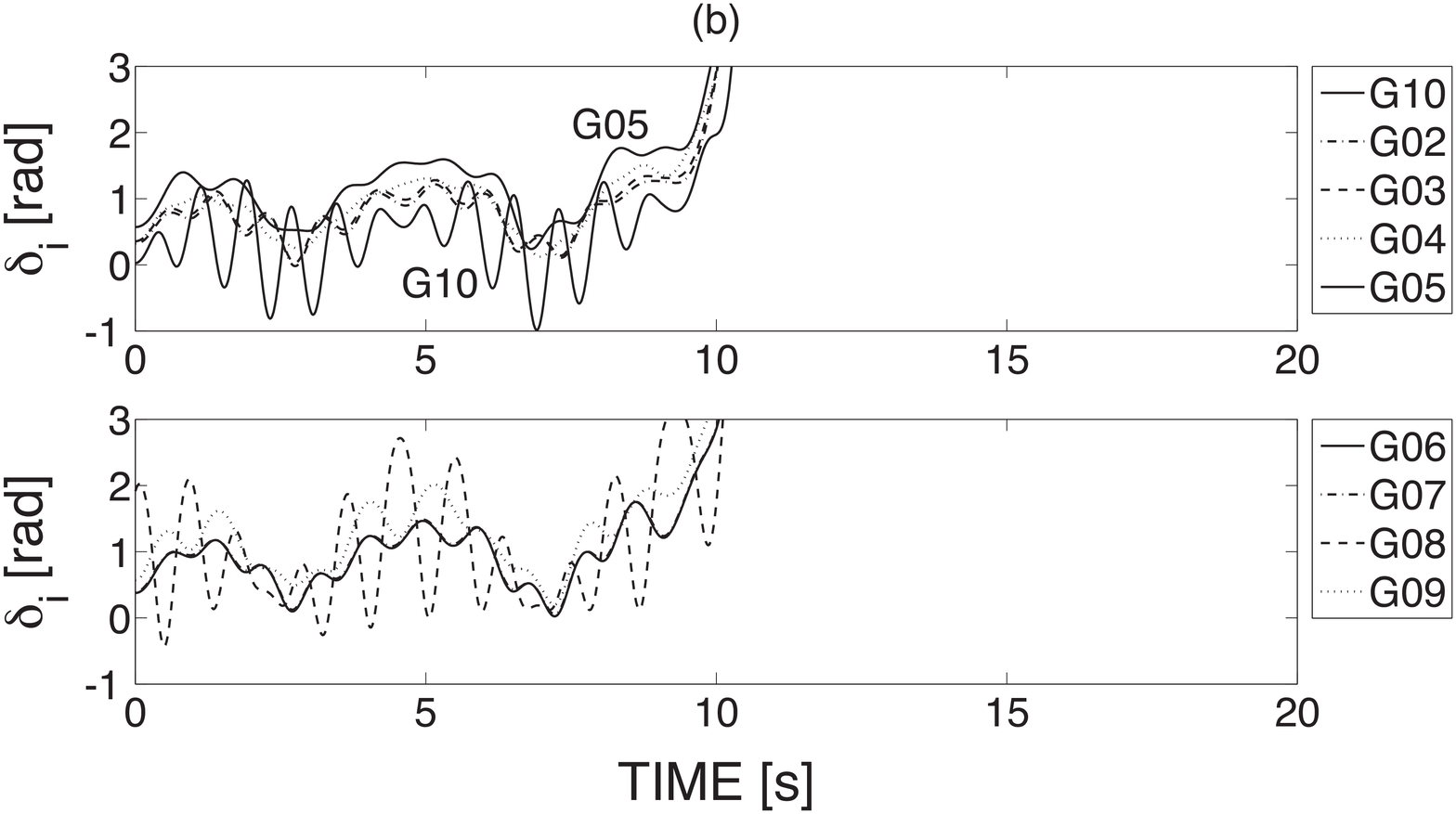}
\caption{%
Coupled swing dynamics and instability observed in the New England 39-bus test system ($D_i=0$):  (a) initial condition (\ref{eqn:ic1}) and (b) initial condition (\ref{eqn:ic2}).  
The angular positions $\delta_i$ only are displayed.  
\copyright 2012 IEEE
}%
\label{fig:waveform2}
\end{figure}

We present other examples of coupled swing dynamics in the NE system.  
Fig.\,\ref{fig:waveform2} shows the time responses of angular positions $\delta_i$ of 9 generators under the two initial conditions:  
\begin{equation}
\makebox[-0.5em]{}(\delta_i(0),\omega_i(0))=
\left\{
\begin{array}{ll}
(\delta^\ast_i+1.000\U{rad},3\U{\,rad/s}) & i=8,
\\\noalign{\vskip +1.0mm}
(\delta^\ast_i,0\U{rad/s}) & \mathrm{else,}
\end{array}
\right.
\label{eqn:ic1}
\end{equation}
and
\begin{equation}
\makebox[-0.5em]{}(\delta_i(0),\omega_i(0))=
\left\{
\begin{array}{ll}
(\delta^\ast_i+1.575\U{rad},3\U{\,rad/s}) & i=8,
\\\noalign{\vskip +1.0mm}
(\delta^\ast_i,0\U{rad/s}) & \mathrm{else.}
\end{array}
\right.
\label{eqn:ic2}
\end{equation}
The initial conditions are close to (\ref{eqn:ic}) and physically correspond to local disturbances at generator 8.  
In Fig.\,\ref{fig:waveform2}(a) the generators do not show divergence motion, that is, they do not show any loss of transient stability for the selected disturbance.    
The behavior is close to that in Fig.\,\ref{fig:waveform1}.  
In Fig.\,\ref{fig:waveform2}(b) they are bounded during the period from $t=0\U{s}$ to $7\U{s}$ and then begin to grow coherently.  
Every generator loses synchronism with the infinite bus at the same time.  
This corresponds to the growth of amplitude of inter-area mode oscillation between the NE system and the infinite bus, namely, the outside of the system.  
This is typical of the CSI phenomenon.   

\begin{figure}[t] 
\begin{center}
\includegraphics[width=0.45\textwidth]{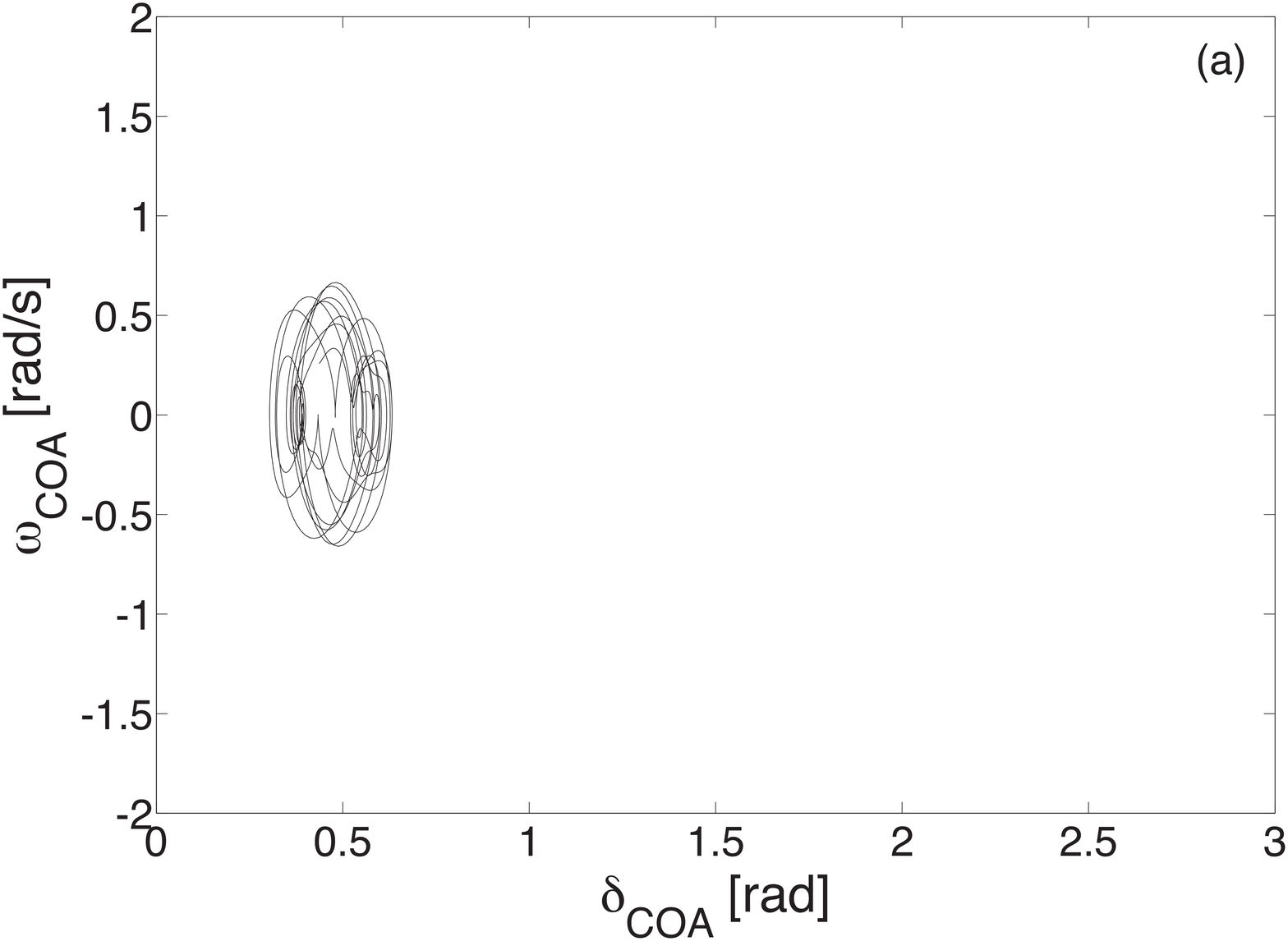} 
\hspace*{4mm}
\includegraphics[width=0.45\textwidth]{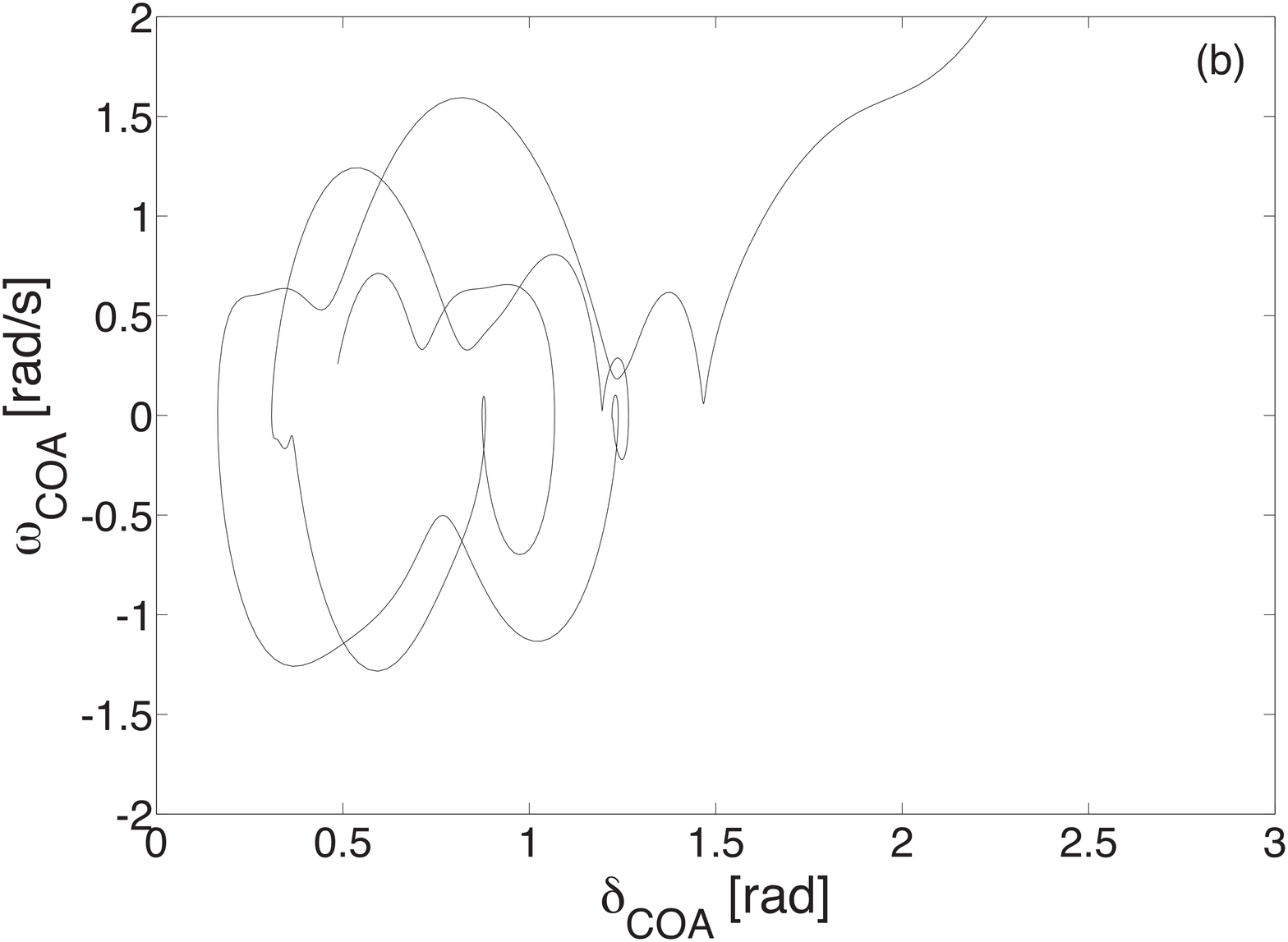}
\caption{%
Collective dynamics in the New England 39-bus test system ($D_i=0$): (a) initial condition (\ref{eqn:ic1}) and (b) initial condition (\ref{eqn:ic2}).  
The Center-Of-Angle (COA) coordinates are defined in (\ref{eqn:COA2}).
\copyright 2012 IEEE
}%
\label{fig:COA2}
\end{center}
\end{figure}

In \cite{Susuki_JNLS09} we showed that CSI involves the divergent motion in the projection of full-system dynamics onto the plane of collective variables.  
The collective variables correspond to the well-known COA (Center-Of-Angle) coordinates in power systems analysis \cite{Athay_IEEE-T-PAS98,Pai:1989}.  
For the NE system, the COA $\delta\sub{COA}$ and its time derivative $\omega\sub{COA}$ are defined as  
\begin{eqnarray}
\delta\sub{COA}:=\sum_{i=2}^{10}\frac{H_i}{H}\delta_i,\makebox[2em]{}
\omega\sub{COA}:=\frac{d\delta\sub{COA}}{dt}=\sum_{i=2}^{10}\frac{H_i}{H}\omega_i,
\label{eqn:COA2}
\end{eqnarray}
where $H:=\sum^{10}_{i=2} H_i$.  
The variables $\delta\sub{COA}$ and $\omega\sub{COA}$ describe the averaged motion of all the generators in the system.  
Fig.\,\ref{fig:COA2} plots the trajectories of (\ref{eqn:classical_NEsystem}) shown in Fig.\,\ref{fig:waveform2} on $\delta\sub{COA}$--$\omega\sub{COA}$ plane.  
The trajectories start near the origin at time $0\U{s}$ and make a couple of almost periodic loops around the initial point.  
In Fig.\,\ref{fig:COA2}(a) the trajectory keeps executing bounded loops.  
In Fig.\,\ref{fig:COA2}(b) it escapes and finally diverges.  

\subsection{Computation of Koopman modes}
\label{subsec:precursor-KMD}

We compute the KMs for the coupled swings shown in Fig.\,\ref{fig:waveform2}(a).  
The computation of KMs is done with the Fourier-based formula (\ref{eqn:projection}).  
The state $\vct{x}$ itself is chosen as the observable, namely $\vct{f}(\vct{x})=\vct{x}$.  
We use the simulation output shown in Fig.\,\ref{fig:waveform2}(a) that extracts the data $\{\vct{x}(nT)\}_{n=0}^{N}$, where the uniform sampling period $T=1/(50\U{Hz})$ and the number of snapshots $N+1=1001$.  
Also, in order to use (\ref{eqn:projection}) for computing KMs, we need to identify dominant frequencies in the course of coupled swings.  
A Fourier analysis shows that the time responses in Fig.\,\ref{fig:waveform2}(a) have clean peaks at the five frequencies, $0.40\U{Hz}$, $0.90\U{Hz}$, $1.00\U{Hz}$, $1.25\U{Hz}$, and $1.45\U{Hz}$.  

\begin{figure}[t] 
\begin{center}
\includegraphics[width=0.5\textwidth]{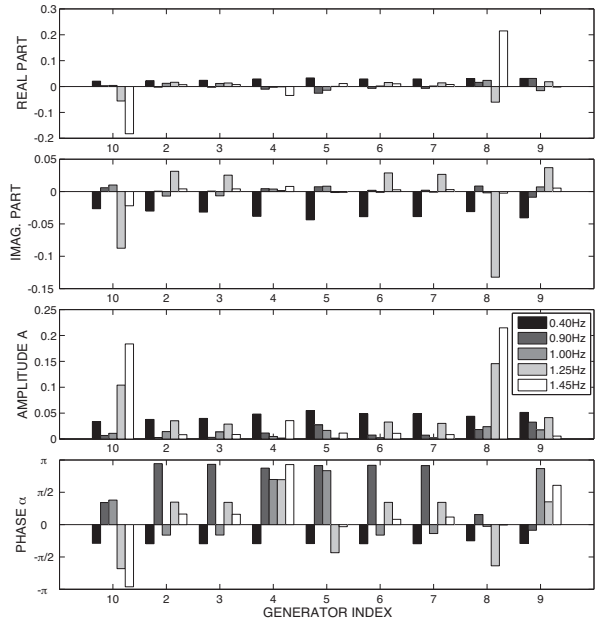}
\caption{%
Computation of Koopman modes ($\phi_i(\vct{x}[0])\vct{V}_i$) using the projection operator $\bP^\nu$ under $\nu=(0.40\U{Hz})T$ at $i=1$, $(0.90\U{Hz})T$ at $i=2$, $(1.00\U{Hz})T$ at $i=3$, $(1.25\U{Hz})T$ at $i=4$, and $(1.45\U{Hz})T$ at $i=5$.  
The symbol $T$ is the sampling period.  
The amplitude coefficients $A_{ji}$ and initial phases $\alpha_{ji}$, defined in (\ref{eqn:te}), are also shown.  
These are obtained for the waveform shown in Fig.\,\ref{fig:waveform2}(a).
\copyright 2012 IEEE
}%
\label{fig:KM2}
\end{center}
\end{figure}

Now we compute the terms $\phi_i(\vct{x}[0])\vct{V}_i$ using the projection operator $\bP^\nu$ with the five frequencies: $i=1$ for 0.40\,Hz, 2 for 0.90\,Hz, 3 for 1.00\,Hz, 4 for 1.25\,Hz, and 5 for 1.45\,Hz.  
The numerical results are shown in Fig.\,\ref{fig:KM2}.  
The angular positions $\delta_i$ only are displayed.  
The amplitude coefficients and initial phases, which are defined in (\ref{eqn:te}), are also shown.  
For 0.40\,Hz, the values of amplitude coefficients are close for each of the generators, and their initial phases are also close.  
All the swings of the 9 generators are hence said to be coherent with respect to the KM with 0.40\,Hz.  
We call it the coherent KM in the same reason as in Section~\ref{sec:coherency}.

\subsection{Precursor diagnostic to coherent swing instability}
\label{subsec:precursor-diagnostic}
We compute the action transfer operator ${\sf J}$ for the CSI phenomenon shown in Fig.\,\ref{fig:waveform2}(b).  
The KMs in Fig.\,\ref{fig:KM2} are computed for the bounded swings in Fig.\,\ref{fig:waveform2}(a).  
The results on Figs.\,\ref{fig:waveform2}(a) and (b) are obtained with different initial conditions.  
Hence, we consider how ${\sf J}$ behaves under dynamics perturbed by a slight change of initial conditions.  
We use the tools (model-order reduction and action-angle representation) developed in Section~\ref{subsec:MOR}.  

\begin{figure}[t] 
\centering
\includegraphics[width=0.7\textwidth]{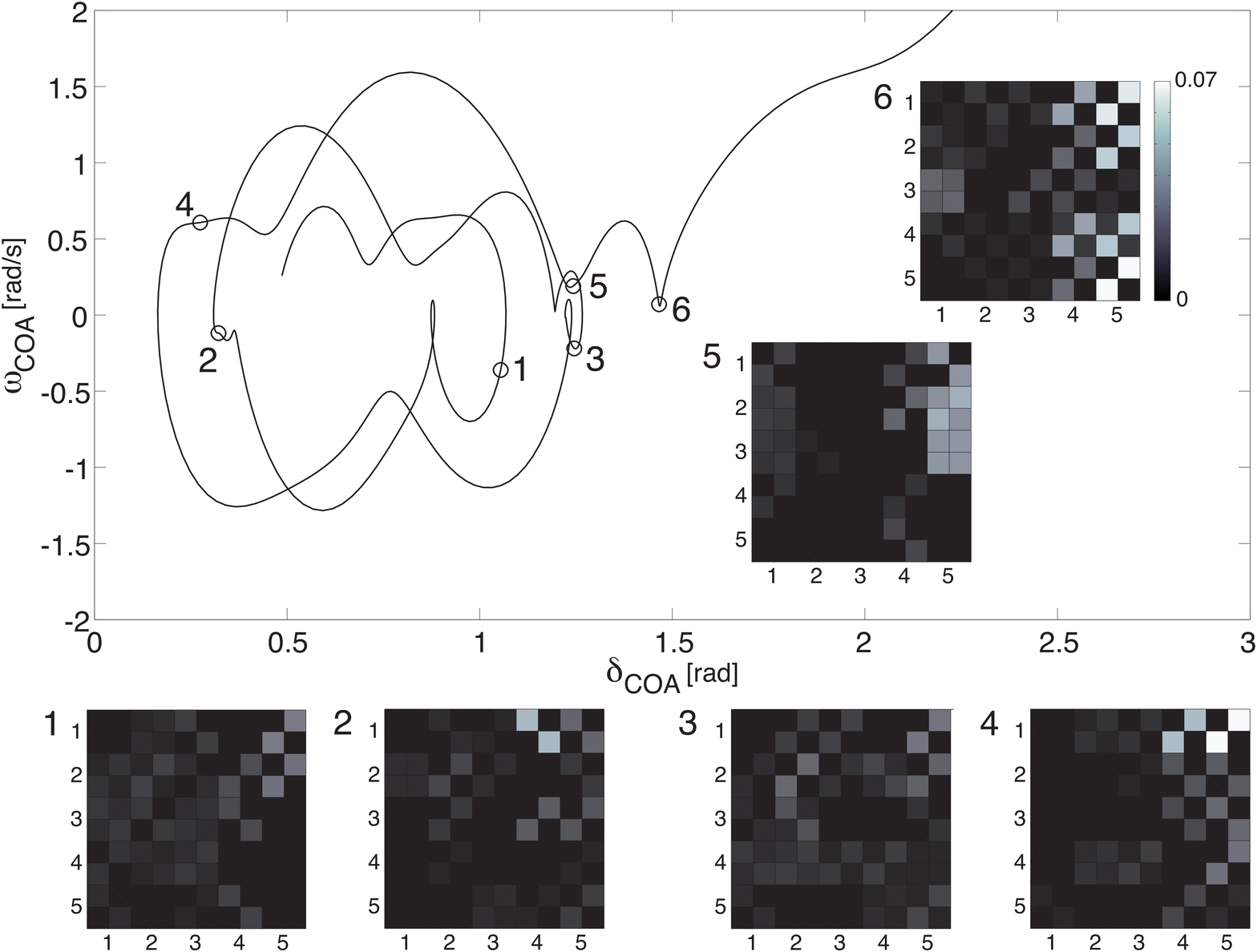}
\caption{%
Collective dynamics of the CSI phenomenon shown in Fig.\,\ref{fig:COA2}(b) and snapshots of the action transfer operator $\mathsf{J}$.  
The six snapshots are at time (1) 1.2\,s, (2) 2.8\,s, (3) 5.0\,s, (4) 7.2\,s, (5) 8.4\,s, and (6) 9.0\,s.
\copyright 2012 IEEE
}%
\label{fig:SS2}
\vspace*{4mm}
\includegraphics[width=0.7\textwidth]{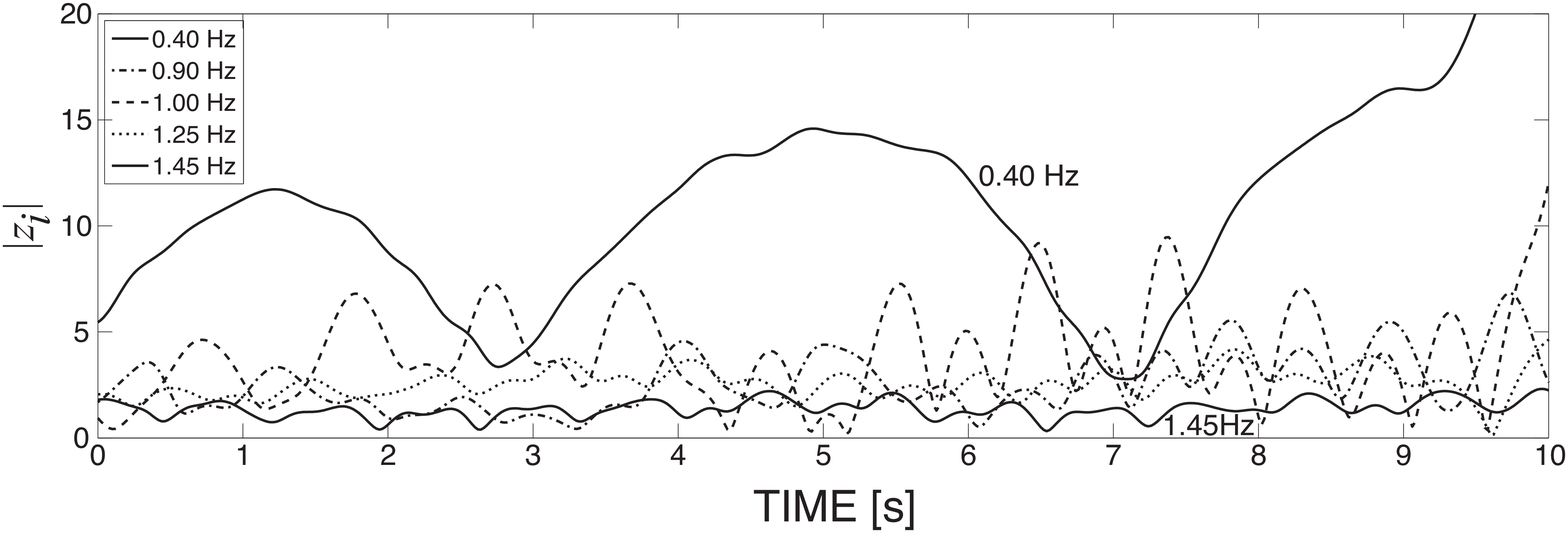}
\caption{%
Time evolution of magnitude $|z_i|$ for the five Koopman modes.  
This is associated with the time evolution of the action transfer operator ${\sf J}$ shown in Fig.\,\ref{fig:SS2}.
\copyright 2012 IEEE
}%
\label{fig:zi2}
\end{figure}

Figure~\ref{fig:SS2} shows several snapshots of the action transfer operator ${\sf J}$ along the COA trajectory shown in Fig.\,\ref{fig:COA2}(b).  
In the snapshots, the $i$-th KM ($i=1,\ldots,5$) corresponds to the $2i-1$ and $2i$ rows.  
The two variables $z_{2i-1}=I_{2i-1}\ee^{\ii\theta_{2i-1}}$ and $z_{2i}=I_{2i}\ee^{\ii\theta_{2i}}$ are assigned to the $i$-th KM.  
By definition, the action variables $I_{2i-1}$ and $I_{2i}$ have the same behaviors, and the angle variables $\theta_{2i-1}$ and $-\theta_{2i}$ also have the same.  
Because both the variables $I_{2i-1}$ and $I_{2i}$ affect the infinitesimal change of $I_j$ ($j=1,\ldots,10$),  in $\mathsf{J}$ we need to plot $\DD \tilde{G}_{j}/\DD I_{2i-1}$ as well as $\DD \tilde{G}_{j}/\DD I_{2i}$.\footnote{See Fig.\,\ref{fig:SS2}.  For the 1st KM ($j=1$), in snapshot~2 the variable $I_{9}$ for the 5th KM dominantly affects the time change of $I_{1}$.  On the other hand, in snapshot~4 the variable $I_{10}$ dominantly affects $I_{1}$.  
These observations both are regarded as the effect of the 5th KM to the 1st KM.  }
First, the magnitude of elements in the upper triangular part change as time goes on.  
On the other hand, the magnitude of elements in the lower triangular part does not change and is almost constant.  
This salient feature of $\mathsf{J}$ suggests a \emph{directed} transmission path of energy from high-frequency KMs to low ones.   
This is consistent with that in \cite{Eisenhower_CDC07} that analyzes global dynamics occurring in nearly-integrable Hamiltonian systems.  
The current result is obtained for a set of KMs embedded in dynamics of a general nonlinear power system model.  
Second, we see that the transmission path of energy from high-frequency KMs to the lowest one becomes stronger when the COA trajectory largely grows in the neighborhood of the $\delta\sub{COA}$-axis: see the snapshots at time (4) 7.2\,s and (6) 9.0\,s in Fig.\,\ref{fig:SS2}.  
The transmission path appears several times, and then the magnitude $|z_1|$ of the lowest KM begins to diverge as shown in Fig.\,\ref{fig:zi2}, while the magnitudes of the other KMs are still bounded.  
Since the lowest KM with 0.40\,Hz is the coherent KM, the divergence  of magnitude $|z_1|$ implies CSI.  
This provides a precursor diagnostic for CSI by a combination of modeling and data: the emergent transmission paths of energy from high-frequency KMs to the lowest coherent KM.

\section{Stability assessment of power systems without models}
\label{sec:stability}

This section provides stability assessment of power systems without any development of models \cite{Susuki_IEEETPWRS29}.  
Many methods for stability assessment of power systems have been developed \cite{Machowski:1997}.  
Traditional methods are mainly model-based, that is, the stability assessment is performed by investigating a mathematical model that represents the target dynamics of a power system.  
It is widely recognized that cascading outages are fairly complicated emergent phenomena in the high-dimensional nonlinear systems.  
Therefore, it is difficult to obtain a mathematical model that explains all events and time evolution of a cascading outage.  
Even if we could obtain such a model, it would be not easy to gain a dynamical insight to the cascading outages from the model because of its complexity.  
Also, the large increase of renewable sources makes it hard to obtain a relevant deterministic model because of their uncertain nature.    
Thus, in contrast to the model-based approach, it is necessary to develop methods that indicate spatio-temporal structure of instabilities and their precursors from data.  
Several reports exist on this line of research \cite{Ostojic_IEEETPS8,Messina_IEEETPS21,Messina_IEEETPS22}.  
In this section, we present an approach to stability assessment based on measured physical power flow data via KMD.  
In contrast to the existing methods, our approach provides not only dynamic patterns of power flows, which we refer to as the base flow patterns below, but also stability information.

\subsection{Mathematical formulation}
\label{subsec:stability-math}

Consider the finite-time data on dynamics of physical power flows (more precisely, active power flows) under uniform sampling, given by
\begin{equation}
\{\vct{P}_0,\vct{P}_1,\ldots,\vct{P}_{N-1}\}, 
\label{eqn:data}
\end{equation}
where $\vct{P}_k\in\bbR^m$ is the snapshot of power flows at the discrete time $k$, $m$ the number of measurement sites (for example, generation plants, transmission lines, and substations), and $N$ the number of available snapshots.  
Generally speaking, the power flows are determined by the internal states of a power system such as rotating frequencies and voltages of AC generators in power plants, bus voltages in substations, power consumptions in loads, and states of controllers in plants and substations \cite{Machowski:1997}.  
Here we use $\vct{x}$ to represent all the internal states belonging to a space $\bbX$ and assume that the time evolution of $\vct{x}$ is represented by the deterministic system as follows:
\[
\vct{x}_{k+1}=\vct{T}(\vct{x}_k),\makebox[2em]{}
k=0,1,\ldots
\]
where $\vct{T}: \bbX\to\bbX$ is a nonlinear vector-valued map and constructed via power system modeling techniques (see, e.g., \cite{Machowski:1997}); thus the dimension of the map is possibly very large.  
Now, in order to associate the finite-time data with the internal state's dynamics, we define a vector-valued observable defined on $\bbX$.  
For the current analysis, it is reasonable to define the observable as a map $\vct{f}_\mathrm{P}: \bbX\to\bbR^m$ such that the snapshot $\vct{P}_k$ of power flows at time $k$ is written as
\[
\vct{P}_k=\vct{f}_\mathrm{P}(\vct{x}_k).
\]

The idea which we now propose for stability assessment is straightforward from KMD.  
Here we develop the analysis for the case when there are unstable eigenvalues of the Koopman operator, but we have access to data for only a short period of time.   
If each of the $m$ components of $\vct{f}_\mathrm{P}$ lies within the span of eigenfunctions $\phi_j$, then vector-valued function $\vct{f}_\mathrm{P}$ is expanded in terms of these eigenfuntions as
\[
\vct{f}_\mathrm{P}(\vct{x})=\sum_{j=1}^{\infty}\phi_j(\vct{x})\vct{V}_j, 
\]
where $\vct{V}_j\in\bbC^m$ are regarded as vector coefficients in the expansion.  
Thus, the time evolution $\vct{P}_k=\vct{f}_\mathrm{P}(\vct{x}_k)$ starting at $\vct{P}_0=\vct{f}_\mathrm{P}(\vct{x}[0])$ is represented as
\begin{equation}
\vct{P}_k=\sum^\infty_{j=1}\lambda^k_j\phi_j(\vct{x}[0])\vct{V}_j,
\label{eqn:infinite}
\end{equation}
where $\vct{V}_j$ is the $j$-th KM associated with KE $\lambda_j$.  
An \emph{unstable} KE is an eigenvalue whose modulus is larger than one, corresponding to the KM that grows exponentially as time increases.  
Here we propose the following method for detection of power system instability: if for a given set of data on power flows obtained over a time interval, there exists no unstable KE, then we conclude that the associated power system is stable.  
If this is not the case, then the power system behaves in an unstable manner.  

Now we discuss the notion of stability addressed in this section.  
Our analysis is based on data for only a short period of time, in other words, is intended to describe \emph{short-term dynamics} exhibited by a nonlinear dynamical system.   
In this sense, the current notion of stability is different from the conventional one that is intended to \emph{long-term, asymptotic} dynamics.  
In a dynamical viewpoint, the current notion deals with a portion of a trajectory and is formulated by examining the unstable eigenvalues of the finite-time Koopman operator corresponding to the time period over which the trajectory is executed.  
The difference of notions of stability can be seen by considering a periodic orbit that slows down as it passes near a saddle and then speeds up.   
Portions of such an orbit could be judged stable or unstable by the method of the section depending on which portion is chosen.  
In this section, we intend to provide a new method for detecting such a short-term trend in dynamics from observational data, thereby obtaining an insight to control measures for (nearly) real-time management.  

We use the above approach to detecting instabilities based on KMD to analyze data on power flows sampled from the 2006 system disturbance of the European interconnected grid.  
Our main idea of the application is to decompose the data on power flows into a set of KMs, implying \emph{base flow patterns}.  
A base flow pattern is geographically distributed over the system, exhibits power flows that grow possibly, and is regarded as a coherent spatial unit of power flows.   
Let $\vct{\tilde{V}}_j=[\tilde{V}_{j1},\tilde{V}_{j2},\ldots,\tilde{V}_{jm}]^\top$ denote the $j$-th empirical Ritz vector associated with the possibly complex-valued Ritz value $\tilde{\lambda}_j=\tilde{r}_j\ee^{\ii 2\pi\tilde{\nu}_j}$ for the finite-time data.  
For convenience we will call the empirical Ritz vector $\vct{\tilde{V}}_j$ the KM, although it differs from it possibly by a multiplicative constant.  
Since the original data are real-valued, if the KM $\vct{\tilde{V}}_j$ is complex-valued, then there exists the conjugate KM $\vct{\tilde{V}}_j^\mathrm{c}$ with the conjugate Ritz value $\tilde{\lambda}_j^\ast$ in the expansion (\ref{eqn:infinite}).  
Here we suppose that the $(j+1)$-th KM $\vct{\tilde{V}}_{j+1}$ corresponds to the conjugate vector $\vct{\tilde{V}}_j^\mathrm{c}$ and the $(j+1)$-th Ritz value $\tilde{\lambda}_{j+1}$ to $\tilde{\lambda}^\ast_j$.  
In this way, the dynamics of base flow pattern described by the KM pair $\{i,i+1\}$ are given in the same manner as (\ref{eqn:te2}):  
\begin{equation}
\vct{P}^{\{j,j+1\}}_k
:=\tilde{\lambda}^k_j\vct{\tilde{V}}_j+(\tilde{\lambda}_{j+1})^k\vct{\tilde{V}}_{j+1}
= 2\tilde{r}^k_j
\begin{bmatrix}
|\tilde{V}_{j1}|\cos\{2\pi k\tilde{\nu}_j+\mathrm{Arg}(\tilde{V}_{j1})\}\\
|\tilde{V}_{j2}|\cos\{2\pi k\tilde{\nu}_j+\mathrm{Arg}(\tilde{V}_{j2})\}\\
\vdots\\
|\tilde{V}_{jm}|\cos\{2\pi k\tilde{\nu}_j+\mathrm{Arg}(\tilde{V}_{jm})\}
\end{bmatrix},
\label{eqn:baseflow}
\end{equation}
where $k=0,1,\ldots,N-2$.  
The spatial shape of the dynamics is captured by the modulus and argument of the KM.  
In the following, we investigate the dynamics of power flows by directly computing the KMs from measured data.

\subsection{Demonstration in 2006 European grid-wide disturbance}
\label{subsec:stability-Europe}

This disturbance was recorded in the night of November 4, 2006 and affected the UCTE's (Union for the Co-ordination of the Transmission of Electricity's) synchronously European interconnected grid \cite{UCTE:2006}.  
It begun in the North German transmission grid, led to a splitting of the UCTE grid into three areas, and caused an interruption of power supply for more than 15 millions European households.  
This splitting is due to the tripping of interconnected AC lines and implies that the three areas were asynchronously operated.  
In each of the areas a significant imbalance of power emerged.  

\begin{figure}[t] 
\centering
\includegraphics[width=0.85\textwidth]{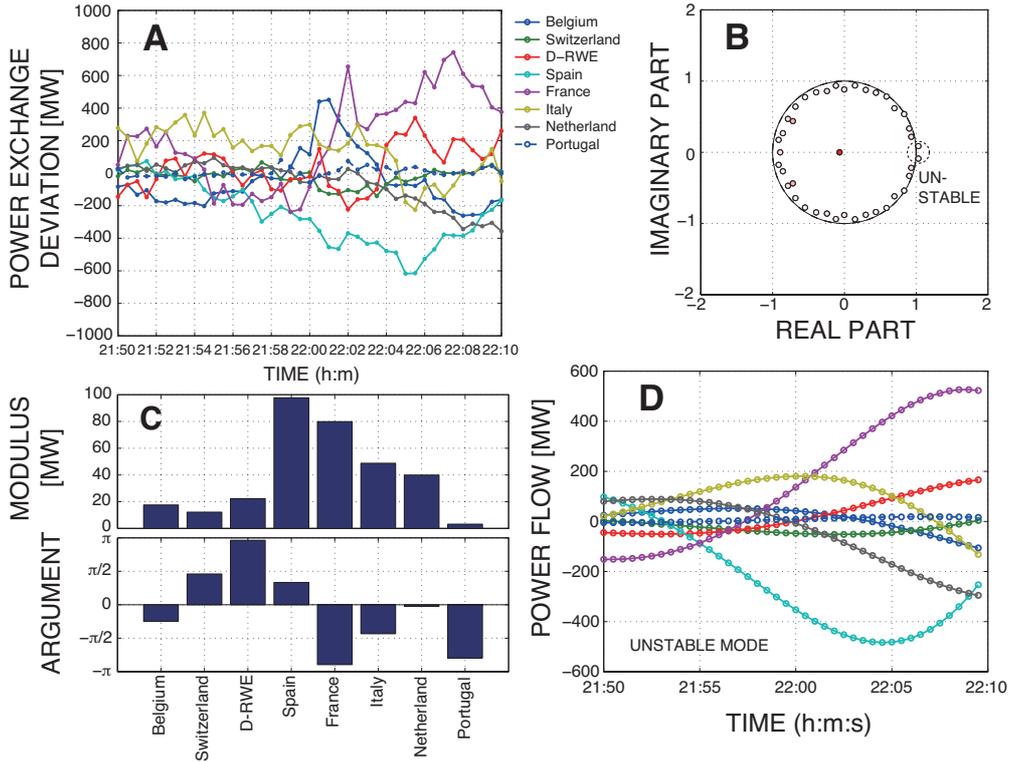}
\caption{%
Koopman mode decomposition of power exchange deviations leading to the 2006 system disturbance in the European grid:  
(A) data on power exchange deviations; 
(B) Koopman eigenvalues $\tilde{\lambda}_j$; 
(C) modulus and argument vectors of unstable Koopman mode; and 
(D) dynamics of base flow patterns (\ref{eqn:baseflow}) for the unstable mode.    
A power exchange deviation is the difference between the actual power exchanges and the scheduled power exchanges with other countries.  
\copyright 2014 IEEE
}%
\label{fig:UCTE}
\end{figure}

The data on dynamics of power flows in the UCTE grid are introduced in Fig.\,\ref{fig:UCTE}A, where we show the dynamics of power exchange deviations from 21:50 till 22:10.  
A power exchange deviation is defined as the difference between the actual power exchanges and the scheduled power exchanges with other countries.  
The areas focused in this figure are Belgium, Switzerland, D-RWE, Spain, France, Italy, Netherland, and Portugal.   
The data in Fig.\,\ref{fig:UCTE}A were obtained by sampling values from the original figure in Appendix~7 of \cite{UCTE:2006} under the uniform sampling (period $T=30\,\mathrm{s}$).  
The final report \cite{UCTE:2006} states, ``These transient deviations are the result of a global shift in physical power flows within the UCTE synchronous grid area to changes in generation programs and exchange programs around 22:00.  
These curves show a situation which is rather normal and typical at that time."  
However, the next single operation (coupling of busbars in the Landesbergen substation in Germany at 22:10) after the transient deviations in Fig.\,\ref{fig:UCTE}A initiated the widespread disturbance in the entire European grid.  
The operation of coupling the busbars was intended to resolve the heavy power flow on the Landesbergen-Wehrendorf 380\,kV line in Germany, but it resulted in its immediate trip and initiated a cascade of line trips, starting with the 220\,kV Bielefeld/Ost-Spexard line and continuing within the E.ON Netz grid \cite{UCTE:2006}.   
Therefore, the N-1 criterion\footnote{The N-1 criterion is a basic principle in power system operation \cite{UCTE:2006}.  This rule is that any single loss of transmission or generation element should not jeopardize the secure operation of the interconnected network, that is, trigger a cascade of line trippings or the loss of a significant amount of consumption.} of secure operation of the grid was not fulfilled before the busbar coupling \cite{UCTE:2006}. 

We performed the KMD of the data on power exchange deviations presented in Fig.\,\ref{fig:UCTE}A.  
The data are the forty-one ($N=41$) snapshots $\{\vct{P}_0,\vct{P}_1,\ldots,\vct{P}_{40}\}$ at eight different $(m=8)$ countries.  
The forty pairs of KEs and KMs $\tilde{\lambda}_j$ and $\vct{\tilde{V}}_j$ were computed and shown in Figs.\,\ref{fig:UCTE}B and \ref{fig:UCTE}C. 
In Fig.\,\ref{fig:UCTE}B, we see that one conjugate pair of $\tilde{\lambda}_j$ exists outside of the unit circle, and the corresponding vector represents the \emph{unstable} mode.  
The other $\tilde{\lambda}_j$'s exist inside of the unit circle, that is, they are associated with the \emph{stable} modes.  
The modulus vectors $[|\tilde{V}_{j1}|,|\tilde{V}_{j2}|,\ldots,|\tilde{V}_{j8}|]^\top$ and argument vectors $[\mathrm{Arg}(\tilde{V}_{j1}),\mathrm{Arg}(\tilde{V}_{j2}),\ldots,\mathrm{Arg}(\tilde{V}_{j8})]^\top$ of the unstable mode are shown in Fig.\,\ref{fig:UCTE}C.  
Also, the dynamics of the unstable base flow pattern are presented in Fig.\,\ref{fig:UCTE}D.  
The period of the unstable mode is 37\,minutes, and thus its oscillation is relatively slow.    
The spatial shape of the mode indicates that there are four large magnitude countries involved.  

\begin{figure}[t] 
\centering
\includegraphics[width=.4\textwidth]{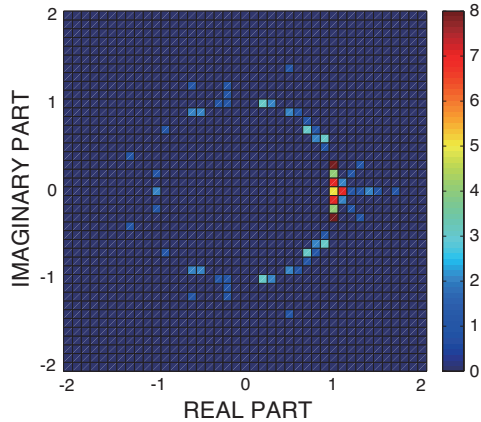}
\caption{%
Number density of unstable Koopman eigenvalues for power exchange deviations leading to the 2006 system disturbance in the European grid 
\copyright 2014 IEEE
}%
\label{fig:UCTE_density}
\end{figure}

The computation shown in Fig.\,\ref{fig:UCTE}B reflects the KEs obtained during the computation over all 41 snapshots.  
It is of interest to consider the evolution of the KE computation over time, to understand the persistence of the unstable KM in relatively noisy data.  
We computed the KEs (and the associated KMs) starting from the 1st sample and computing until $N=8,9,\ldots,40$, namely, for many sets of snapshots with different final number $N$.  
The result is shown in Fig.\,\ref{fig:UCTE_density} as the number density plot, where the color changes from \emph{blue} to \emph{black} and depicts the number density of unstable KEs in each cell. 
In this figure, the \emph{black} cells are close to the unstable KEs shown in Fig.\,\ref{fig:UCTE}.   
Therefore, the unstable KM persists under a change of the sample length.  
As reviewed above, the dynamics of power exchange deviations are closely related to the initiation of the system disturbance.  
The base flow pattern of the unstable KM presented in Fig.\,\ref{fig:UCTE}D vividly shows how the large-scale unstable dynamics can be extracted from the noisy data in the report---the signature of the instability, in other words, the breaking of the N-1 criterion is clearly present in the dynamics even before time 22:00.  
It should be recalled that the UCTE report argues that the situation is rather normal and typical at that time.

\section{Conclusion and future problems}
\label{sec:outro}

In this paper, we have outlined our recent efforts on applications of the Koopman operator theory to power systems technology.  
The first part of this paper (Section~\ref{sec:theory}) was devoted to an introduction to the Koopman operator theory for nonlinear dynamical systems, with emphasis on modal decomposition and computation.  
The second part (Sections~\ref{sec:coherency} to \ref{sec:stability}) was devoted to the three topics of the applications:  coherency identification of swings in coupled synchronous generators, precursor diagnostic of instabilities in the coupled swing dynamics, and stability assessment of power systems without any development of models.  
We emphasize that these are established as data-centric methods, namely, how to use massive quantities of data obtained numerically and experimentally, via the KMD (Koopman Mode Decomposition).  

Here we note a topic excluded due to space constraints of this paper.  
In \cite{Raak_IEEETPWRS2015,Raak_IFACCPES15} the authors demonstrate the KMD's capability for partitioning a power network, which determines the points of separation in an islanding strategy.  
A practical data-driven method incorporating the KMD for network partitioning is proposed.  
It is shown that the data-driven method can cover network partitions derived from both spectral graph theory \cite{Fiedler75} and slow-coherency-based method \cite{Yusof_IEEETPS8}.  
The KMD-based partitioning pins down the frequency of oscillations as well as information on damping and participation, thus it can be used for monitoring and control purposes, namely the controlled islanding technique: see, e.g., \cite{YouVittal2004}.

In closing, we identify two future problems, not addressed in or following the current analysis in Sections~\ref{sec:coherency} to \ref{sec:stability}: 
\begin{itemize}
\item The applications presented here are specific to swing dynamics governing active power transmission and frequency maintenance.  
The power system has  the other important physical aspect: reactive power and voltage.  
Loss of the normal level of voltage is often crucial in the course of cascading outages \cite{Kurita_CDC88,Sweden:2003,NorthAmerica:2003}.  
This poses a problem of characterization of \emph{non-oscillatory} responses in terms of spectrum of the Koopman operator, which is still a difficult problem in the existing Koopman operator techniques.   
\item Control and operation of the power system in terms of the Koopman operator should be pursued.  
As stated in the beginning of Section~\ref{sec:intro}, the dynamics of power systems occur in a wide range of scales in space and time.  
Depending on our purposes of control and operation, we need to extract dynamical information from the data collected practically or predicted in numerical simulations, and then to make a decision or actuation in order to archive a desirable physical response.  
KMD has two main advantages:  (i) clear separation of spatio-temporal scales, even from relatively-noisy data as demonstrated in Section~\ref{sec:stability}, and (ii) direct computation from data without any use of description of the underling system.
\end{itemize}

\section*{Acknowledgements}

The first author appreciates Dr.\,Ryan Mohr, Mr.\,Hassan Arbabi, and Prof.\,Alexandre Mauroy for valuable discussions. 
He also thanks many collaborators and students, especially, Prof.\,Yutaka Ota, Mr.\,Yohei Kono, and Mr.\,Kyoichi Sako for valuable discussion on individual theoretical and applied topics. 
The work presented in this paper is supported in part by JSPS Postdoctoral Fellowships for Research Abroad, JST-CREST \#JPMJCR15K3, and MEXT KAKEN \#15H03964.


\end{document}